\documentclass[a4paper,twoside,10pt]{article}
\usepackage{color}
\usepackage{figsize}
\usepackage[left=2cm,right=2cm,top=2cm,bottom=2cm]{geometry}
\usepackage{natbib}
\usepackage{lineno}

\begin{document}

\begin{center}
	\textbf{\LARGE Radiative-convective models of the atmospheres of Uranus and Neptune: heating sources \& seasonal effects}~\\[1cm]
	\normalsize Gwenael Milcareck$^{a,b}$, Sandrine Guerlet$^{a,c}$, Franck Montmessin$^{b}$, Aymeric Spiga$^{a}$, Jeremy Leconte$^{d}$, Ehouarn Millour$^{a}$, Noe Clement$^d$, Leigh N. Fletcher$^{e}$, Michael T. Roman$^{e}$, Emmanuel Lellouch$^{c}$, Raphael Moreno$^{c}$, Thibault Cavalie$^{d,c}$, Oscar Carrion-Gonzalez$^{c}$ ~\\[0.2cm]
	\small $^a$ \emph{Laboratoire de Météorologie Dynamique/Institut Pierre-Simon Laplace (LMD/IPSL), Sorbonne Université, CNRS, \'{E}cole Polytechnique, Institut Polytechnique de Paris, \'{E}cole Normale Supérieure (ENS), PSL Research University, 4 place Jussieu BC99, 75005 Paris, France}\\
	\small \emph{$^b$ Laboratoire Atmosphères, Milieux, Observations spatiales (LATMOS), IPSL, Observatoire de Versailles St-Quentin-en-Yvelines, Université de Versailles St-Quentin-en-Yvelines, CNRS, 11 boulevard d’Alembert, 78280 Guyancourt, France}\\
	\small \emph{$^c$ Laboratoire d’Etudes Spatiales et d’Instrumentation en Astrophysique (LESIA), Observatoire de Paris, CNRS, Sorbonne Université, Université Paris-Diderot, Meudon, France}\\
	\small \emph{$^d$ University of Bordeaux, CNRS, LAB, UMR 5804, Pessac, France}\\
	\small \emph{$^e$ School of Physics \& Astronomy, University of Leicester, University Road, Leicester, LE1 7RH, United Kingdom}
\end{center}

\begin{center}
	\rule{4cm}{0.4mm}
\end{center}

\begin{abstract}
   The observations made during the Voyager~2 flyby have shown that the stratosphere of Uranus and Neptune are warmer than expected by previous models. In addition, no seasonal variability of the thermal structure has been observed on Uranus since Voyager~2 era and significant subseasonal variations have been revealed on Neptune.
   In this paper, we evaluate different realistic heat sources that can induce sufficient heating to warm the atmosphere of these planets and we estimate the seasonal effects on the thermal structure. 
   The seasonal radiative-convective model developed by the Laboratoire de Météorologie Dynamique is used to reproduce the thermal structure of these planets. Three hypotheses for the heating sources are explored separately: aerosol layers, a higher methane mole fraction, and thermospheric conduction.
   Our modelling indicates that aerosols with plausible scattering properties can produce the requisite heating for Uranus, but not for Neptune. Alternatively, greater stratospheric methane abundances can provide the missing heating on both planets, but the large values needed are inconsistent with current observational constraints. In contrast, adding thermospheric conduction cannot warm alone the stratosphere of both planets. The combination of these heat sources is also investigated. In the upper troposphere of both planets, the meridional thermal structures produced by our model are found inconsistent with those retrieved from Voyager~2/IRIS data. Furthermore, our models predict seasonal variations should exist within the stratospheres of both planets while observations showed that Uranus seems to be invariant to meridional contrasts and only subseasonal temperature trends are visible on Neptune. However, a warm south pole is seen in our simulations of Neptune as observed since 2003.
\end{abstract}


\newpage

\section{Introduction}

Located at 19 and 30 AU respectively from the Sun, Uranus and Neptune are known as cold worlds. Indeed, the irradiance received at the top of their atmospheres is particularly low: it is 3.69 W.m$^{-2}$ for Uranus and 1.51 W.m$^{-2}$ for Neptune (on Earth, it is 1361 W.m$^{-2}$). With an albedo of 0.30 and 0.29 for Uranus \citep{Pearl1990} and Neptune \citep{Pearl1991} respectively, the energy balance implies an absorbed energy flux of 0.64 W.m$^{-2}$ for Uranus and 0.27 W.m$^ {-2}$ for Neptune. These planets differ in their internal energy flux: it is $0.042 \pm 0.047$ W.m$^{-2}$ at most for Uranus \citep{Pearl1990} while it is estimated at $0.43 \pm 0.09 $ W.m$^{-2}$ on Neptune \citep{Pearl1991}. Thus, the internal energy flux is greater than the absorbed solar energy in the case of Neptune. With an obliquity of $28.32^\circ$, seasonal variations are expected on Neptune. In the case of Uranus, its obliquity of $97.77^\circ$ means that its rotation axis is almost on its orbital plane and thus, on an annual average, Uranus receives a greater solar flux at the poles than at the equator, unlike the other planets of the solar system. In summary, these two planets are characterised by low sunlight and long orbital periods ($\sim$84 terrestrial years for Uranus and $\sim$165 years for Neptune), by very marked seasonal variations in irradiance (especially for Uranus), and, in the case of Neptune, strong competition between internal energy flux and absorbed solar flux. These differences impact the energy balance of these planets and therefore their atmospheric temperatures.~\\

Temperature measurements are notably difficult due to the distance and the low infrared radiation emitted by these planets. The most precise measurements of the upper tropospheric temperature structure come from the Voyager~2 flyby. The radio-occultation experiment provided 1-D profiles on Uranus at latitude 2-6° S during northern winter solstice at $\sim271^\circ$ in solar longitude (Ls) \citep{Lindal1987} and at latitude 59-62° N during northern autumn (Ls$\simeq$235°) on Neptune \citep{Lindal1992} (see fig.~\ref{fig:noaerosol}). At 1000 hPa, the temperatures reach $\sim$76 K on Uranus and $\sim$71 K on Neptune.
Temperature profiles derived from disk-averaged spectra from Spitzer Infrared Spectrometer for Uranus \citep{Orton2014}, Infrared Space Observatory for Neptune \citep{Burgdorf2003} and AKARI \citep{Fletcher2010} confirm the temperature observed in the troposphere by Voyager~2 (fig.~\ref{fig:noaerosol}). 

Near 100 hPa, a particularly cold tropopause has been observed by Voyager~2. From radio-occultations, the minimum is 53 K on Uranus and 52 K for Neptune. These temperatures are cold enough for methane to condense. IRIS observations from Voyager~2 reveal a complex meridional thermal structure near the tropopause for Uranus \citep{Flasar1987,Conrath1998,Orton2015} and Neptune \citep{Conrath1989,Conrath1991,Conrath1998,Fletcher2014}. On zonal-mean temperatures maps, an equatorial maximum and a local minimum at mid-latitudes in each hemisphere are observed for both planets. 

The stratosphere of both planets has been observed by Voyager~2, space-based and ground-based observatories. Radio-occultations from Voyager~2 indicate a temperature of the order of $\sim$80 K at 1~hPa on Uranus and $\sim$125 K at the same level on Neptune. The Voyager~2 PPS experiment provided information on temperatures in the uranian lower-stratosphere near 1~hPa at 68.9°N \citep{Lane1986,West1987} through a UV stellar occultation. Assuming an aerosol-free atmosphere, the PPS temperature retrieval showed a lower stratosphere warmer by 10 K at $\sim$3~hPa than temperature measurements from radio-occultation. \citet{Greathouse2011} and \citet{Fletcher2014} explored meridional temperature variations in Neptune's stratosphere with thermal-infrared images from Keck/LWS (2003), Gemini-N/MICHELLE (2005), VLT/VISIR (2006), Gemini-S/TReCS (2007) and Gemini-N/TEXES (2007). Assuming a spatially constant methane abundance, the stratospheric temperature seems to be latitudinally-isothermal since the Voyager~2 flyby. However, \citet{Roman2022} showed important spatial and temporal variations in the meridional temperature.
On Uranus, \citet{Roman2020} suggest that the meridional temperature gradient displays a similar structure to that seen at the tropopause. In the upper stratosphere, UVS measurements from Voyager~2 and ground-based stellar occultations showed that this region is particularly hot (on average 150 K at 1 Pa for both planets). However, observations from Voyager~2 and from the Earth are inconsistent for both planets. Temperatures measured from the Earth are lower than those from Voyager~2 and vary very strongly vertically. Temperature differences can reach 100 K at 1 Pa \citep{Saunders2023}.


To interpret the observed temperature profiles, several radiative-convective equilibrium models have been built and used. The simulated stratospheric temperatures are, by far, too cold compared to the observed ones. The temperature mismatch can reach 30 K in the lower-stratosphere for both planets \citep{Wallace1983,Appleby1986,Friedson1987,Marley1999, Greathouse2011, Li2018}. This "energy crisis" \citep{Friedson1987} is also present in the thermosphere where there are several hundred kelvins differences between observations and models \citep{Melin2019}. Many hypotheses have been explored to explain this difference in the stratosphere and thermosphere. \citet{Appleby1986} concluded that the presence of a "continuum absorber" in the stratosphere -- which may be aerosols -- could contribute significantly to the energy balance on Uranus but not entirely on Neptune. However, \citet{Marley1999} showed that adding stratospheric hazes based on the assumption of spherical Mie scattering particles did not warm this region appreciably on Uranus. 
On Neptune, the same conclusion was established \citep{Moses1995}. Alternatively, a heat flux from an unknown source in the thermosphere \citep{Stevens1993} which conducts heat to lower levels is not sufficient to warm the stratosphere of Uranus \citep{Marley1999} and Neptune \citep{Wang1993}. On Uranus, by adding a small abundance of methane in the lower thermosphere, \citet{Marley1999} managed to reconcile the observed and modelled temperatures, as in this scenario the stratosphere is warmed by the methane that radiates downward \citep{Marley1999}. However, this scenario implies elevated methane abundance values at homopause levels that are inconsistent with Voyager~2 and ground-based observations.

Previous radiative-convective models predicted that Uranus' stratospheric temperatures should undergo seasonal variations over the course of its 84-year orbital period \citep{Wallace1983,Friedson1987,Conrath1990}. Nonetheless, no similar trend has been observed since the Voyager~2 flyby \citep{Roman2020} except on stellar occultations where an apparent seasonal variation is possible in the high stratosphere \citep{Young2001,Hammel2006}. On the contrary, the temperature has changed considerably on Neptune since Voyager~2 era. Significant sub-seasonal variations in the stratosphere have been discovered that could be related to solar activity \citep{Roman2022} or inertia-gravity waves \citep{Hammel2006,Uckert2014}. At higher pressures, models predict limited seasonal effects on temperature near the tropopause \citep{Wallace1984}, and the thermal structure seems to have remained invariant since Voyager~2 flyby \citep{Orton2015,Roman2020}, except at the poles \citep{Fletcher2014}.


Understanding the origin of the "energy crisis" on the ice giants is one of the current major challenges in planetary science. Current 1-D radiative-convective models do not reproduce the observed temperature structure or the seasonal variability without adding one or more heating sources. 
To reproduce the thermal structure of the atmosphere of Uranus and Neptune and its seasonal variability, a seasonal radiative-convective model designed for ice giants is introduced in section \ref{sec:methodology} with a description of different parameters used. The temperature profile obtained with our nominal model is described in section \ref{sec:initial-profile}. Section \ref{sec:heat-sources} explores several additional heat sources that can warm the stratosphere of both planets. The simulated meridional temperature structure is discussed and compared with the observations in section \ref{sec:seasonal}, along with the expected seasonal variability.

\section{Methodology \label{sec:methodology}}

Here, we describe the 1-D seasonal radiative-convective equilibrium model tailored for ice giants developed at Laboratoire de Météorologie Dynamique (LMD), which aims at understanding the radiative heat sources, the seasonal variability and the thermal structure of Uranus' and Neptune's atmospheres. This model was previously used to simulate the radiative forcing on exoplanets \citep{Wordsworth2010,Turbet2016} and, more recently, on Jupiter and Saturn \citep{Guerlet2014,Guerlet2020}. 

Radiative transfer equations are solved in a column of atmosphere discretized in layers using the two-stream approximation, including multiple scattering as proposed by \citet{Toon1989} and depending on opacity sources that control the heating and cooling rates. In addition to these opacity sources, a radiative flux at the bottom corresponding to the measured internal heat flux is also added in the case of Neptune (not for Uranus, as it is negligible). As suggested by \citet{Zhang2023a,Zhang2023b}, the internal heat flux in ice giants could vary with latitude and even fluctuate over time, however, this effect has never been quantified and we did not explore this possibility in our study. To emulate convective mixing, a convective adjustment scheme relaxes the temperature profile towards the adiabatic lapse rate \citep{Hourdin1993} if an unstable lapse rate is encountered during a simulation. The tropospheric lapse rate is controlled by the standard gravity and the heat capacity fixed at one value. We choose to fix the heat capacity at the value calculated at 3000 hPa (corresponds to the bottom of our model) by using the temperature observed and the abundance of hydrogen for a given ortho:para ratio, helium and methane at this level. On Uranus, a heat capacity of 8600 J.K$^{-1}$.kg$^{-1}$ is found, consistent with the "intermediate" ortho:para ratio case from \citet{Massie1982}. On Neptune, the heat capacity is chosen for H$_2$ at equilibrium and set at 9100 J.K$^{-1}$.kg$^{-1}$.

Orbital and planetary settings are added to take seasonal effects on temperature into account. The most important parameter is the obliquity, which is set to 97.77° for Uranus and 28.32° for Neptune. Because Uranus and Neptune have radiative time constants ranging from years to decades \citep{Conrath1990,Conrath1998,Li2018}, a daily-averaged solar flux is considered and calculations of the radiative heating and cooling rates are performed typically once every 25 planetary days.

All seasonal radiative-convective simulations presented in this paper employ a pressure grid consisting of 48 levels between 3000 hPa and 5 Pa which covers the upper troposphere and the lower stratosphere. The radiative spin-up is ensured by running 30 Uranian years and 16 Neptunian years. 
Regarding the radiative forcings, all opacity contributions are separated into two parts: a thermal infrared one (10-3200 cm$^{-1}$) which controls the cooling rate and a visible one (2020-33300 cm$^{-1}$) which controls the heating rate. Like gas giant planets, radiative cooling results from collision-induced absorption (mainly H$_2$–H$_2$ and H$_2$–He) in the thermal infrared in the lower atmosphere along with thermal emission by the main hydrocarbons (CH$_4$, C$_2$H$_2$, C$_2$H$_6$) in the stratosphere. Radiative heating results from the absorption of visible and near-infrared solar photons by methane and collision-induced absorption in the lower atmosphere \citep{Conrath1990,Conrath1998,Li2018}. The effect of clouds and hazes are considered later in section~\ref{sec:heat-sources-aerosols}.

\begin{figure}[ht!]
    \centering
    \includegraphics[width=.45\textheight]{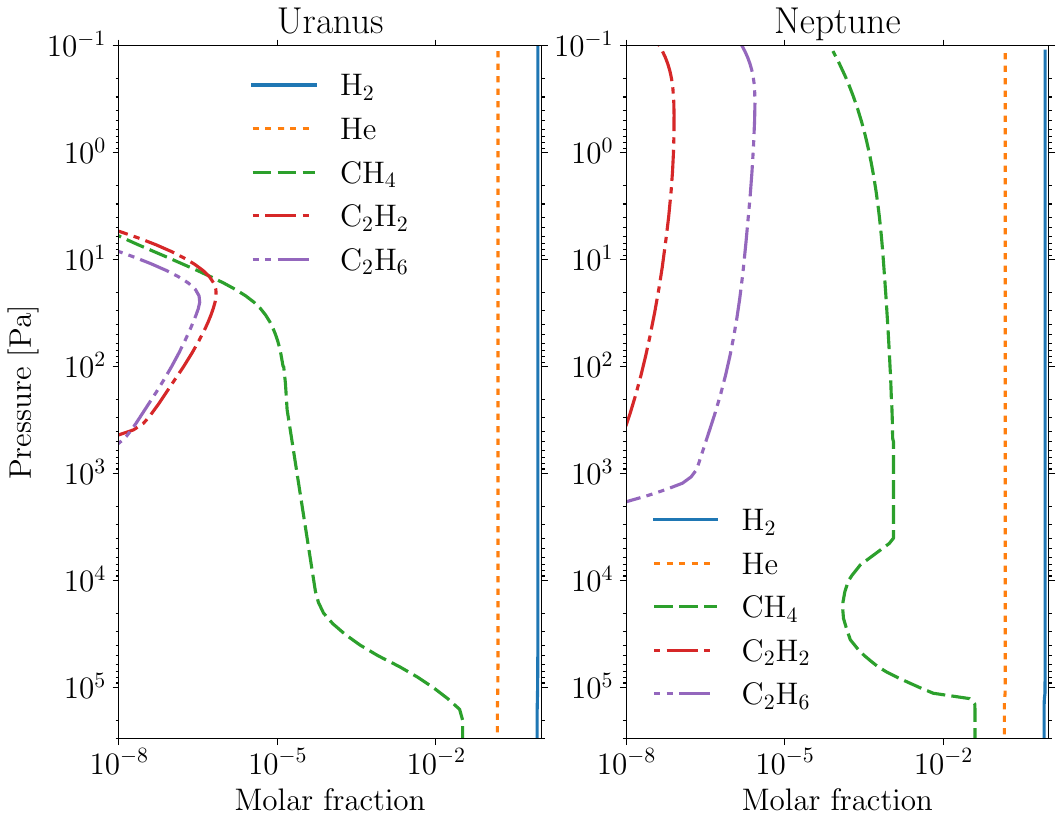}
    \caption{\label{fig:resume_compo} Vertical molar fraction of the gases on Uranus (left) and Neptune (right) used by the model. Collision-induced absorptions/emissions of H2 and He contribute to radiative heating/cooling throughout the entire atmospheric column, as absorption/emission by methane (except on Uranus, where its abundance becomes insignificant at pressures below $\sim$10 Pa). The other hydrocarbons are active only in the stratosphere for pressures between $\sim$1 hPa and $\sim$7 Pa on Uranus and below than $\sim$10 hPa on Neptune.}
\end{figure}

Thermal emission and visible/near-infrared absorption by hydrocarbons are key to the radiative cooling and heating in ice giant atmospheres. As line-by-line calculations are too time-consuming for model applications, correlated-k coefficients for different spectral bands and temperature–pressure values are pre-calculated offline \citep{Goody1989,Wordsworth2010}. The k-distribution model was obtained as described by \citet{Guerlet2014} for Saturn with the same compounds.
Briefly, we start by computing the high-resolution spectra of the absorption coefficients $k(\nu)$ of hydrocarbons (CH$_4$, C$_2$H$_2$, C$_2$H$_6$) line-by-line from the spectroscopic data of HITRAN 2016 and \citet{Rey2018} according to their vertical distribution (fig. \ref{fig:resume_compo}) and their methane isotope content (CH$_3$D, $^{13}$C) on a rough temperature-pressure grid (9 values between 40 and 190 K and 12 layers between 10$^6$ and 0.1 Pa).  Then, the spectra are discretized into several spectral bands (20 bands for the thermal part and 26 bands in the visible part) between 0.3 and 1000 $\mu$m which results from a good compromise between bandwidth and number of bands in order to maintain good precision and a fairly short reading time. For each spectral band and T-p value, high-resolution absorption coefficients $k(\nu)$ are sorted by strength, then converted into a cumulative probability function $g(k)$ (with $g$ varying from 0 to 1), and finally this function is inverted to obtain $k(g)$. This is now a smooth function that replaces $k(\nu)$ and can be integrated over typically 16 Gauss points (8 sampled between 0 and 0.95, plus 8 sampled between 0.95 and 1). Like \citet{Guerlet2014}, H$_2$-broadened coefficients are used instead of air-broadened coefficients for methane and ethane \citep{Halsey1988,Margolis1993}. During a GCM run, these tabulated coefficients are interpolated at the temperature computed by the model at a given time step, at each  pressure level. Non-local thermodynamic equilibrium effects from CH$_4$ on the temperature are not taken into account because it starts to be significant at pressures lower than 3 Pa for Uranus and 0.5 Pa for Neptune \citep{Appleby1990}, which is above the top boundary of our model. Note that all opacity sources (not only the hydrocarbon ones) are calculated on the spectral bands used by the k-distribution model.

We consider collision-induced absorption (CIA) from H$_2$-H$_2$ \citep{Borysow1991, Zheng1995, Borysow2000, Fletcher2018}, H$_2$-He \citep{Borysow1989a, Borysow1989b}, H$_2$-CH$_4$ \citep{Borysow1986}, He-CH$_4$ \citep{Taylor1988} and CH$_4$-CH$_4$ \citep{Borysow1987} according to the vertical distribution of each species (fig. \ref{fig:resume_compo}). For H$_2$-H$_2$, H$_2$-He and H$_2$-CH$_4$ data, the hydrogen ortho-para ratio is set to the equilibrium as observed (on average) on both planets.

Rayleigh scattering following the method described in \citet{Hansen1974} from the three main gases (H$_2$, He, CH$_4$) is included. Raman scattering by molecular hydrogen is neglected because its optical depth is lower that of Rayleigh scattering \citep{Sromovsky2005} and the heating/cooling rate of Rayleigh scattering is already much lower than hydrocarbons or CIA contributions (100 to 1000 times lower). CIA and Rayleigh scattering are interpolated on the same grid of the k-distribution model (pressure, temperature and wavelength).

The abundance of methane on Uranus and Neptune is known to vary with latitude and altitude. Indeed, in the troposphere of Neptune, the molar fraction of methane at the equator reaches 6 to 8\% while at the poles, it decreases to 2 to 4\% \citep{Karkoschka2011,Irwin2019b}. On Uranus, the gradient is less marked with a methane mole fraction reaching 3 to 4\% at the equator and $\sim$1\% at the poles \citep{Karkoschka2009,Sromovsky2019}.
This latitudinal variation is accompanied by a vertical variation linked to the condensation of methane near 1000 hPa and to the photochemistry, eddy and molecular diffusion in the middle and upper atmosphere. 
On Uranus, due to a more stratified atmosphere, the methane homopause exists at higher pressures (between 10 and 1 Pa) than on Neptune (between 10$^{-2}$ and 10$^{-3}$ Pa) \citep{Moses2018}. These variations have the consequence of strongly playing on the competition between the different sources of opacity according to latitude and altitude. The detailed examination of methane's vertical profile can be found in Section~\ref{sec:heat-sources-methane}. In the case of the model used here, only variations in altitude are taken into account and the methane is assumed to be horizontally uniform and constant over time. For Uranus, the methane mole fraction profile below the 100 Pa level (3.2\%) from \citet{Lellouch2015} is combined with the annual average profile from the photochemical model of \citet{Moses2018} above the 100 Pa level (fig.~\ref{fig:resume_compo}). On Neptune, the methane deep tropospheric value is set to 4\% based on latitudinally averaged retrieved values by \citet{Irwin2019b}; near the condensation level, the profile from \citet{Lellouch2015} is used and above this level, the annual average profile from \citet{Moses2018} is chosen. We take this variation on the abundance of H$_2$ and He into account, and the molar fraction of He/H$_2$ is fixed at 0.15/0.85 for both planets \citep{Conrath1987,Burgdorf2003}. Concerning the vertical distribution of hydrocarbons (C$_2$H$_2$,  C$_2$H$_6$, CH$_3$D, $^{13}$C), they are set to an annual-averaged profile derived from the photochemical model of \citet{Moses2018}. Variations in chemical distributions are predicted by photochemical models which alter the heating/cooling rates but the effect is expected to be secondary to direct variation in seasonal insolation at the pressures considered.

\section{Vertical thermal structure: clear-sky models \label{sec:initial-profile}}

\begin{figure}[ht!]
    \centering
    \includegraphics[width=.45\textheight]{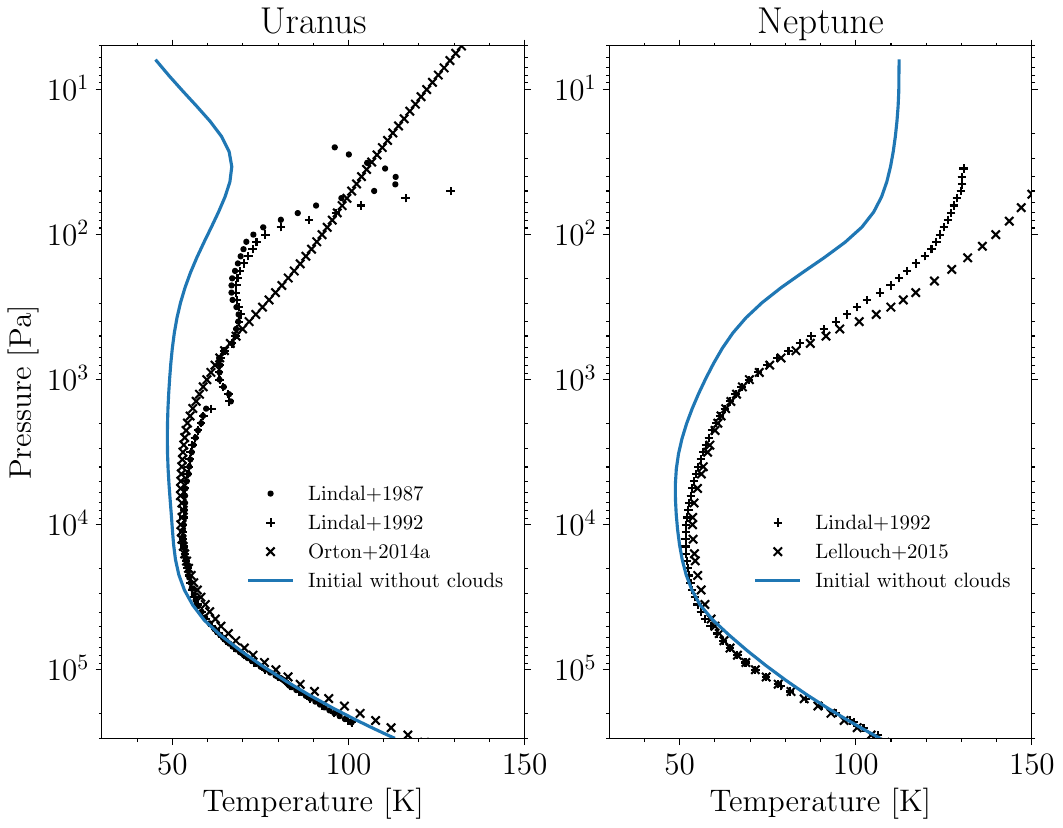}
    \caption{\label{fig:noaerosol} Temperature simulated by the nominal model in blue on Uranus (left) and Neptune (right) at the same latitude and solar longitude as the observations (see text). Observed temperature profiles  are from \citet{Lindal1987} (black dots), \citet{Lindal1992} (black plus), \citet{Orton2014} (black cross) for Uranus and \citet{Lellouch2015} (black cross) for Neptune.}
\end{figure}

Using our 1-D radiative-convective model with the parameters described above, we simulated temperature profiles for both planets at the latitudes and solar longitudes corresponding to Voyager~2 radio-occultation profiles. The temperature profiles that derived from Voyager~2 were obtained at 2--6°S and Ls$\simeq$271° for Uranus  \citep{Lindal1987,Lindal1992} and at 59--62°N and Ls$\simeq$235° for Neptune \citep{Lindal1992}. We caution that the \citet{Lindal1992} Voyager~2 temperature profile of Neptune was derived assuming a He/H2 ratio of 0.19/0.81 that is higher than the more recently derived 0.15/0.85 ratio \citep{Burgdorf2003}. Using a lower He/H2 ratio in the radio-occultation data analysis would lead to a temperature profile colder by only a few Kelvins, and would not change our conclusions. To complete data in the stratosphere for comparison purposes, the globally-averaged temperatures retrieved by \citet{Orton2014} at Ls$\simeq$0° for Uranus and \citet{Lellouch2015} at Ls$\simeq$279° for Neptune are taking into account.
The simulated temperature profiles obtained here are shown in fig.~\ref{fig:noaerosol}. 

The predicted tropospheric temperatures for both planets are similar as expected. This similarity is due to the lack of internal heat flux on Uranus and to the excess in the case of Neptune. Sensitivity tests were performed by parameterising the internal energy flux at the upper and lower limits estimated by \citet{Pearl1990,Pearl1991}. On Uranus, the tropospheric temperature obtained at 3000 hPa is warmer by 2 K and on Neptune, it is 5 K warmer (resp. colder) using the high (resp. lower) limit of 0.53 W.m$^{-2}$ (resp. 0.38 W.m$^{-2}$). The tropospheric temperatures on Neptune are thus sensitive to the internal heat flux.
Concerning their stratosphere, as in previous studies, an important gap called "energy crisis" exists between the temperature retrieved by the observations and the simulated ones. The difference begins at the top of the troposphere on Uranus ($\sim$500 hPa) and at the tropopause on Neptune ($\sim$100 hPa). The temperature mismatch reaches $\sim$70 K on Uranus at 10 Pa and $\sim$25 K on Neptune at 100 Pa. Our model predicts also that Neptune's stratosphere is warmer than Uranus' despite its larger distance to the Sun. For instance, at 1 hPa, our model predicts that Neptune is 35 K warmer than Uranus.
This can be explained by a significantly higher abundance of methane in Neptune's stratosphere (1.3$\times$10$^{-5}$ on Uranus and 1$\times$10$^{-4}$ on Neptune at 1 hPa), implying a larger heating rate. This is consistent with the results of \citet{Li2018} who computed the different contributions of the gaseous opacity sources to the heating and cooling rates: at pressures lower than 50~hPa, methane is the dominant heating source. 
Regarding the cooling rates, on Uranus, they are dominated by the CIA opacities throughout the atmosphere while on Neptune, CIA opacities control the cooling rates at pressures higher than 1--2 hPa and hydrocarbons are the dominant contributors at lower pressures. 
We note that like in our simulations, \citet{Li2018} (who did not take haze opacities into account) reported a heating deficit in both planets' stratospheres. 

To further evaluate our results, we computed the Bond albedo, based on simulations performed at all latitudes.
\citet{Pearl1990} and \citet{Pearl1991} inferred a Bond albedo of $0.30\pm0.05$ and $0.29\pm0.05$ based on Voyager~2 observations for Uranus and Neptune, respectively. Our values of 0.27 and 0.25 are in rough agreement with the ones derived from observations. We note however that the reanalysis of full-disk reflectivity data from Voyager~2 flyby hints at a Bond albedo that may be higher on Uranus and lower on Neptune \citep{Wenkert2022}.

In the next section, we explore other radiative heat sources that could increase the temperature above the tropopause while maintaining a realistic Bond albedo value (keeping in mind that these values may be overestimated/underestimated). 

\section{Supplementary heat sources in the stratosphere \label{sec:heat-sources}}

\subsection{Aerosol and cloud layers \label{sec:heat-sources-aerosols}}

Thermochemical models \citep{Pryor1992,Baines1994} predict, from the temperature profile and the hydrocarbon abundances (CH$_4$, C$_2$H$_6$, C$_2$H$_2$), that a thick methane cloud should be located at $\sim$1500 hPa and that other hydrocarbons should condense in the lower stratosphere and be optically thin for both planets. Several scenarios of the vertical distribution of clouds and hazes have been proposed for Uranus and Neptune to reproduce observations in different portions of the visible and near-infrared (NIR) spectrum.
However, numerous sources of uncertainty exist that make the characterisation of the haze/cloud structure very challenging. These include uncertainties in the spectral dependence of optical properties, in the latitudinal and vertical variation of methane abundance, in seasonal changes of cloud distribution, in the difficulty in seeing the contribution of haze/cloud opacities due to the spectral dominance of methane gas in the NIR and uncertainties linked to the narrow spectral windows of observations.
Nevertheless, these studies have established a first-order vertical structure on ice giants. 

In the domain of study of our model (between 3000 hPa and 5 Pa), Uranus' atmosphere is thought to comprise at least one optically thin haze layer with a particle mean radius of $\sim$0.1~$\mu$m located above the CH$_4$ condensation level and an optically thick cloud layer located between 1000 and 3000 hPa with larger particle size ($\sim$1~$\mu$m). The existence of the methane cloud layer remains although thermodynamically expected because some scenarios do not need such a cloud at the level of methane condensation \citep{Sromovsky2007, Karkoschka2009, Irwin2012b, Irwin2015, Roman2018, Sromovsky2019}.
Some more complex haze/cloud scenarios with more haze layers exist \citep{Sromovsky2007,Sromovsky2011,Sromovsky2014}. Concerning the optical properties of the haze located in the upper troposphere \citep{Tice2013, Sromovsky2014} or the lower stratosphere \citep{Sromovsky2007,Karkoschka2009,Sromovsky2011,Tice2013,Irwin2015,Irwin2017}, they are poorly constrained. Moreover, latitudinal variations in optical depth, refractive index and particle radius have been found recently for the upper haze \citep{Sromovsky2011, Roman2018, Sromovsky2019}, adding difficulty in identifying the optical properties of haze particles.


Concerning Neptune, a lot of scenarios have also been proposed to reproduce the vertical haze/cloud structure.
These studies retrieved a similar vertical haze/cloud structure but optically thinner in comparison to Uranus. The haze particle radii are also found to be smaller ($\sim$0.1 $\mu$m) than that of the deep cloud particles ($\sim$1 $\mu$m) \citep{Karkoschka2011,Cook2016,Irwin2019b}. As on Uranus, the optical properties are also poorly constrained \citep{Karkoschka2011,Irwin2011a,Irwin2016b,Cook2016,Irwin2019b} and the existence and altitude of the tropospheric methane cloud is uncertain \citep{Karkoschka2011}. Due to the intense cloud activity, the retrieved properties from various scenarios are also  latitudinally dependent.

\begin{figure}[h]
        \centering
        \includegraphics[width=.45\textheight]{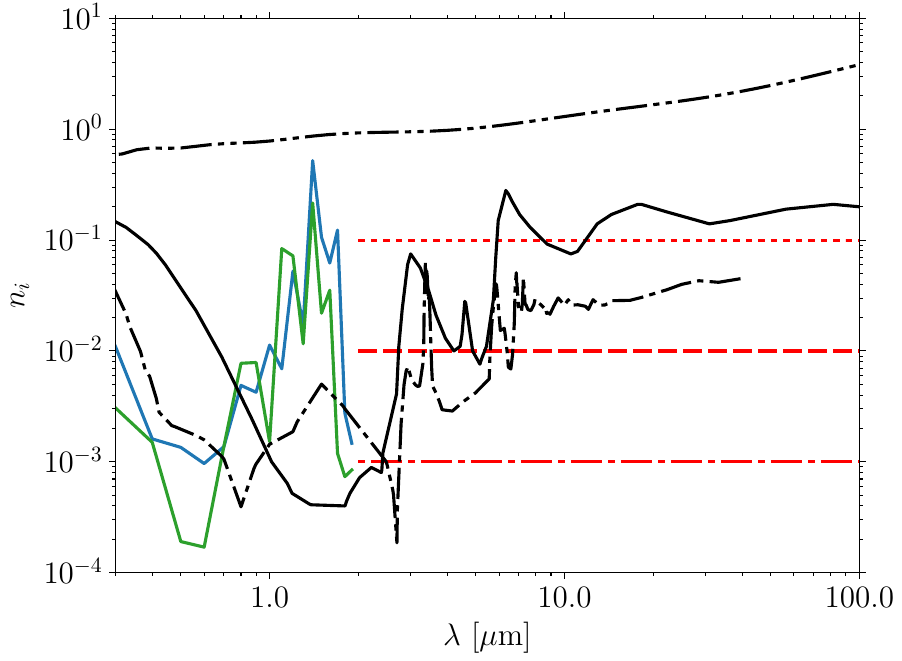}
        \caption{\label{fig:imaginary_uranus}Uranus: Imaginary refractive indices for Uranus. The blue line corresponds to the vertically extended photochemical haze and the green line to the concentrated haze located near the CH$_4$ condensation level from \citet{Irwin2022}. Three different ad hoc values (1$\times$10$^{-1}$, 1$\times$10$^{-2}$, 1$\times$10$^{-3}$) in the thermal infrared (>2 $\mu$m) are added (red lines). For reference, refractive indices of tholins \citep{Khare1984}, ice tholins \citep{Khare1993} and black carbon \citep{Jager1998} are also shown in solid, dotted-dashed and double dotted-dashed black lines respectively.}
\end{figure}

\begin{figure}[h]
        \centering
        \includegraphics[width=.45\textheight]{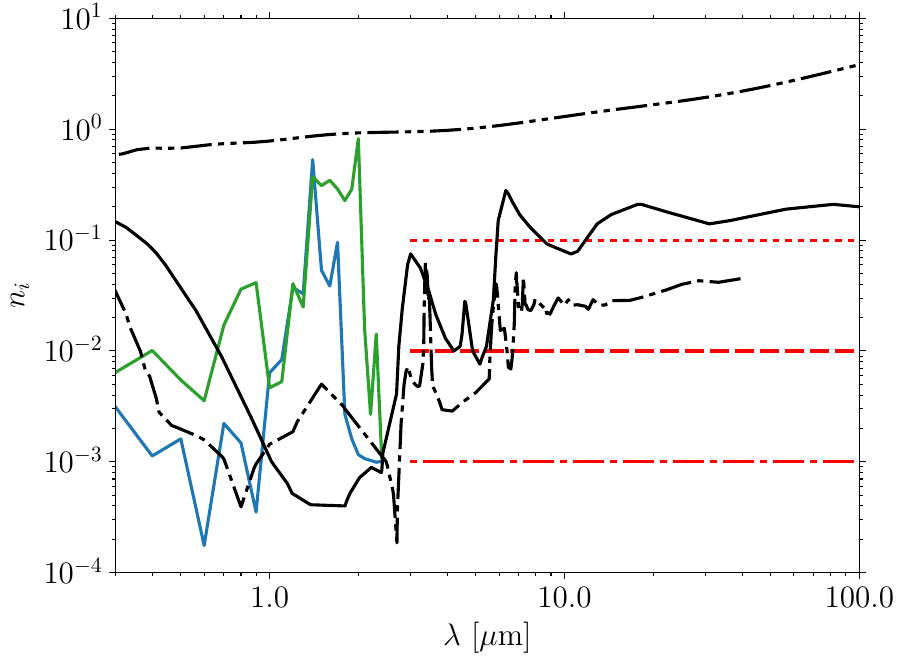}
        \caption{\label{fig:imaginary_neptune}Same as fig. \ref{fig:imaginary_uranus} but for Neptune.}
\end{figure}

Nearly all these models were based on narrow-band observations, with a limited spectral range. \citet{Irwin2022} reanalysed a combined set of observations (IRTF/SpeX, HST/STIS, Gemini/NIFS) to cover a broader spectral range (0.3--2.5~$\mu$m) than other studies in order to better characterise the vertical structure of haze/cloud layers in ice giants and their optical properties from visible to near-infrared light. By using this combined spectrum and a radiative transfer model, the following vertical structure (in the vertical range of our model) has been found for both planets:
\begin{itemize}
    \item A vertically extended photochemical haze at pressures lower than 1000 hPa composed of submicron-sized particles which are more scattering at visible wavelengths and more absorbing at UV and longer wavelengths (see their imaginary refractive index in fig.\ref{fig:imaginary_uranus} for Uranus and fig.\ref{fig:imaginary_neptune} for Neptune); 
    \item A compact aerosol layer concentrated near the methane condensation level at 1000-2000 hPa composed of micron-sized particles with similar optical properties to the optically-thin extended haze above it (see fig.\ref{fig:imaginary_uranus} and \ref{fig:imaginary_neptune}).
\end{itemize}

On Neptune, \citet{Irwin2022} added a thin methane cloud layer near the tropopause to fit the reflectance spectrum at wavelengths longer than $\sim$1 $\mu$m. We note that this tropopause cloud is very spatially variable rather than globally uniform, corresponding to the patchy clouds seen in Near IR data. Concerning the absence of methane cloud at $\sim$1000 hPa on Uranus and Neptune, the authors argue that the presence of aerosols at 1000-2000 hPa acting as cloud-condensation nuclei, causes such a rapid condensation of methane that the newly formed methane ice precipitates instantly. Because the study by \citet{Irwin2022} is the most comprehensive one to date, we thus base our aerosol parametrisation on their results.

In our model, the effects of clouds and hazes on radiative heating/cooling rates are simulated using the following inputs: their total optical depth, their vertical distribution (parameterised with a given bottom pressure level and a fractional scale height), the weighted mean of their particle radius distribution with their effective variance, and the optical properties of the species. The latter are the extinction coefficient, the single scattering albedo and the asymmetry factor which are calculated by a Mie code \citep{Bohren1983} as a function of wavelength for a given particle radius and refractive index (the real part $n_r$ and the imaginary part $n_i$). 
The vertical distribution of the vertically extended photochemical haze on both planets is less constrained. For the next investigations, we adopt the value of the fractional scale height adopted by \citet{Irwin2022} where the aerosols are uniformly present from 1600 hPa to the top of our model ($\sim$5 Pa). The reality is necessarily more complex, with a more irregular distribution and an altitude above which the atmosphere is effectively clear of hazes.

\begin{table*}[ht]
  \centering
  \begin{tabular}{ c | l | c | c }
     Cloud/haze layer & Parameter & Uranus & Neptune \\ \hline
     Vertically & Bottom pressure (hPa) $p_1$ & 1600 & 1600 \\
     extended & Fractional scale height $f_1$ & 2.0 & 2.0 \\
     photochemical & Optical depth (at 0.8 $\mu$m) $\tau_1$ & 0.04 & 0.05 \\
     haze & Particle radius ($\mu$m) $r_1$ & 0.05 & 0.05 \\ \hline
     Haze & Bottom pressure (hPa) $p_2$ & 1500 & 2000 \\
     concentrated & Fractional scale height $f_2$ & 0.1 & 0.1 \\
     near CH$_4$& Optical depth (at 0.8 $\mu$m) $\tau_2$ & 2.0 & 1.0 \\
     condensation level & Particle radius ($\mu$m) $r_2$ & 0.8 & 0.8 \\ \hline
     Methane & Bottom pressure (hPa) $p_{CH_4}$ & - & 200 \\
     cloud & Fractional scale height $f_{CH_4}$ & - & 0.1 \\
     at the & Optical depth (at 0.8 $\mu$m) $\tau_{CH_4}$ & - & 0.03 \\
     tropopause & Particle radius ($\mu$m) $r_{CH_4}$ & - & 2.5 \\ \hline
  \end{tabular}
  \caption{\label{tab:irwin2022} Best-fit haze scenario adapted from \citet{Irwin2022} that is consistent with the Bond albedo retrieved during Voyager~2 flyby \citep{Pearl1990,Pearl1991}.}
\end{table*}

For the hazes, the weighted average of all best-fitting retrieved refractive index ($n_i$) spectra deduced by \citet{Irwin2022} are used, but they are only available in the visible/NIR part of the spectrum. Knowing the refractive index in the thermal infrared is necessary to account for thermal emission from hazes. Thus, three different ad hoc values of refractive index (1$\times$10$^{-1}$, 1$\times$10$^{-2}$, 1$\times$10$^{-3}$) in this spectral range have been tested (fig.\ref{fig:imaginary_uranus} for Uranus and fig.\ref{fig:imaginary_neptune} for Neptune) to estimate the amount of radiative cooling by these hazes. In the case of Neptune, the refractive index from \citet{Martonchik1994} is used for the additional methane cloud located at the tropopause.

Before evaluating the radiative impact of aerosols on the temperature profile, preliminary tests according to the ranges of the different parameters of these cloud and haze layers (Table 2 in \citet{Irwin2022}) are necessary in order to verify if the Bond albedo obtained by our model is consistent with the one retrieved during the Voyager~2 flyby. The best-fit scenario for both planets is given in table \ref{tab:irwin2022} and the Bond albedo from these parameters is 0.35 for Uranus and 0.34 for Neptune, which corresponds for both planets to the upper limit of the Bond albedo retrieved during Voyager~2 era. Knowing that the Bond albedo observed is maybe overestimated on Neptune, the albedo obtained by our model may not be realistic. We note that without the methane cloud layer only at the tropopause, the Bond albedo obtained is equal to 0.29.

\begin{figure}[h]
    \centering
    \includegraphics[width=.45\textheight]{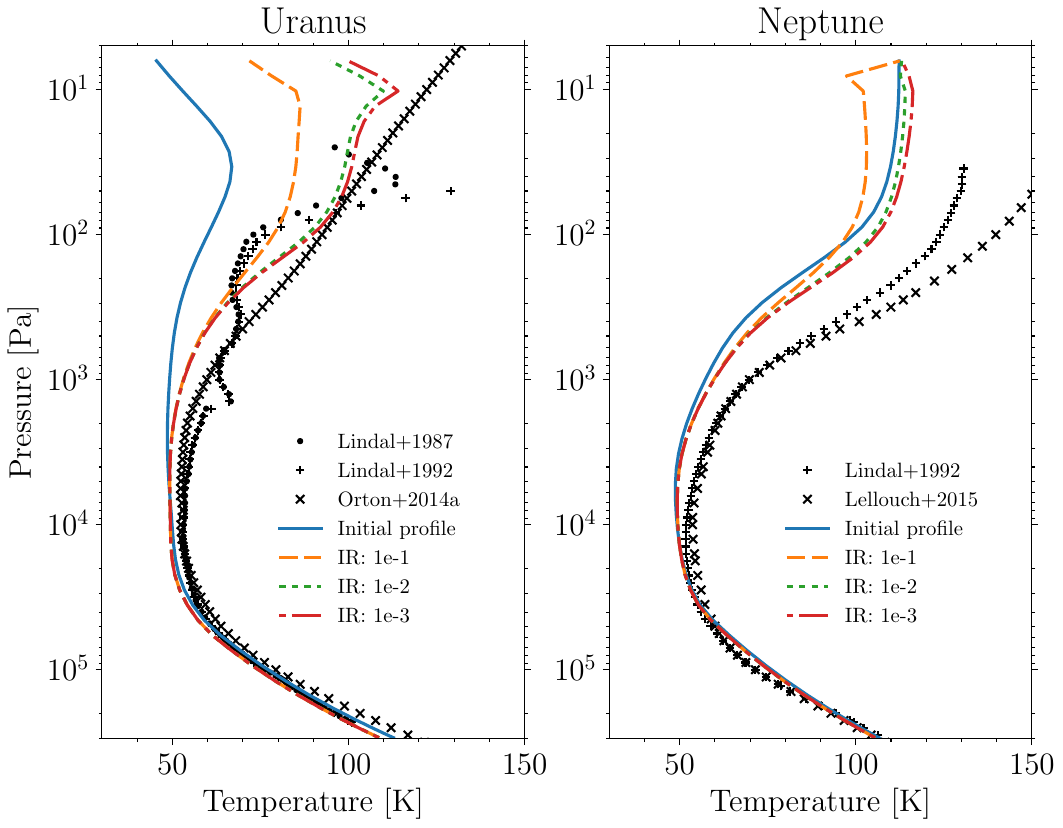}
    \caption{\label{fig:irwin2022}Simulated temperature profiles of Uranus (left) and Neptune (right) with the \citet{Irwin2022} haze scenario for three different values of the absorption coefficient in the thermal infrared. The orange dashed line is the case with an imaginary index of 10$^{-1}$ in the thermal infrared, the green dotted line is the 10$^{-2}$ case and the red dotted-dashed line is the 10$^{-3}$ case. For $n_i$ lower than 10$^{-3}$ in the thermal infrared, we obtain the same temperature profile as the case at 10$^{-3}$. In comparison, the temperature reached without aerosols is added as a blue line and the different observations are shown in black as described in fig.\ref{fig:noaerosol}.}
\end{figure}

\begin{figure}[h]
    \centering
    \includegraphics[width=.45\textheight]{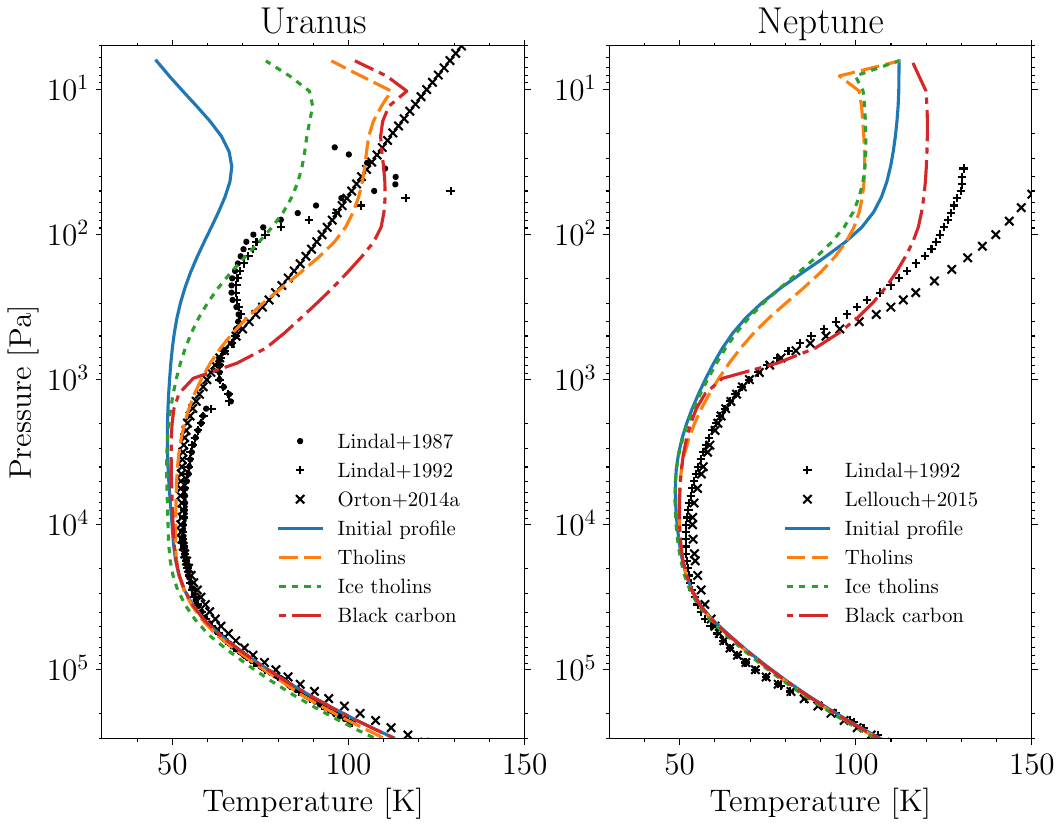}
    \caption{\label{fig:tholins}Simulated temperature profiles of Uranus (left) and Neptune (right) with the \citet{Irwin2022} haze scenario but with different optical indices. The orange dashed line is the case where the \citet{Irwin2022} optical indices are replaced by those of tholins \citep{Khare1984} and the green dotted line corresponds to ice tholins \citep{Khare1993}. The red dotted-dashed line is the temperature simulated with an adhoc black carbon dust layer located between 5 and 1000 Pa. In comparison, the temperature reached without aerosols is displayed as a blue line and the different observations are shown in black as described in fig.\ref{fig:noaerosol}.}
\end{figure}

We find that including hazes significantly warms Uranus' stratosphere: our simulated temperature profiles for the three different thermal infrared refractive index values closely approximate Voyager~2 observations (fig.\ref{fig:irwin2022}). The difference between the simulated temperature without aerosol layers and the observed temperature reached 70 K in the lower stratosphere ($\sim$10 Pa). By adding these aerosols, it is only $\sim$5-10 K at this level depending on the value of $n_i$ in the thermal infrared. At the tropopause and lower stratosphere, the thermal structure is now more consistent with the observed tropopause level than the case without haze layers. This difference is explained by the absence of radiative species sufficiently absorbent to warm the atmosphere at these levels. The refractive index of this aerosol layer is fairly high near the ultraviolet and in the near-infrared, and the opacity is high enough for the absorption to be significant to allow heating. The results of our simulations are different from previous publications where it was assumed that aerosols had little or no effect on heating the stratosphere \citep{Marley1999,Moses1995}. This will be discussed in section \ref{sec:heat-sources-discussion}. 
However, at pressures below 30 Pa, the heating is no longer sufficient to maintain a profile similar to that observed, due to a lower opacity of the aerosol layer. The profile becomes almost isothermal and departs from the temperature profile observed on Uranus. Another heating source seems to be required at the top of our model to better match the observations.
Complementary tests by changing the optical indices of aerosol layers by those of tholins and ice tholins were performed. 
Ice tholins-like particles indeed seem to be a good candidate for the haze \citep{Irwin2022} as they have similar refractive indexes (Fig.\ref{fig:imaginary_uranus} \& \ref{fig:imaginary_neptune}). When using the ice tholins optical indices (Fig. \ref{fig:tholins}), we also report a warming effect but it is less important than when using the haze properties  of \citet{Irwin2022}. Rather, a similar warming is simulated by our model by replacing the optical indices of \citet{Irwin2022} by those of tholins.  

In the case of Neptune, the heating produced by aerosol layers is in a relative sense less important than the one obtained on Uranus (Fig.\ref{fig:irwin2022}) because absorption in near-infrared light is already dominated by methane in the stratosphere (unlike Uranus). Adding the aerosol scenario only adds 5 K compared to the simulation without aerosol at $\sim$1~hPa. Moreover, assuming the highest value for $n_i$ in the thermal infrared yields a net cooling effect for p$<$1~hPa due to a large thermal emission. 
At pressures above the 1~hPa level, the temperature becomes also almost isothermal. By replacing the optical indices by tholins or ice tholins, no significant change is observed (Fig. \ref{fig:tholins}). 

Another interesting candidate for the haze material is the rings of these planets. UVS observations showed that the hydrogen exosphere of Uranus extends to the rings and therefore can transport dust materials into the atmosphere \citep{Broadfoot1986,Herbert1987}. \citet{Rizk1990} showed that dust particles falling from rings in a small latitude band centred at the equator can significantly warm the high stratosphere of Uranus. The particles from the rings are known to be very dark, similar to black carbon \citep{Ockert1987,Karkoschka1997}. We performed tests by adding an arbitrary haze layer with optical constants of black carbon retrieved in laboratory from a pyrolysis experiment \citep{Jager1998} without the haze scenario from \citet{Irwin2022}. An optically thin ($\tau=0.01$ at 160 nm) layer of this type of particle with a small radius (0.1 $\mu$m) confined arbitrarily between 1000 and 5 Pa can warm this entire vertical range. The Bond albedo of 0.27 obtained is consistent with the value calculated during the Voyager~2 era (0.30$\pm$5). Adding this layer on Neptune gives a similar warming (fig.\ref{fig:tholins}) but the optical depth must be greater than on Uranus ($\tau=0.05$ at 160 nm). This greater opacity means that the atmosphere is darker, with a Bond albedo equal to 0.22, inconsistent with the observed value (0.29$\pm$5). Moreover, such a ring-flowing material on Neptune remains speculative due to the lack of observations of an exosphere extending to the rings.

Similarly to \citet{West1991} and \citet{Marley1999}, we find that ethane (C$_2$H$_6$), acetylene (C$_2$H$_2$) and diacetylene (C$_4$H$_2$) ices are insufficiently dark in the UV and visible and too optically thin to absorb significant amounts of solar flux on both planets by using \citet{Baines1994} haze scenario. Acetylene haze has an anti-greenhouse effect and results in a decrease in temperature by 5 to 10 K on Uranus. This haze could be responsible for the low temperature observed between 70 and 300 Pa. On Neptune, no significant effect is visible.

\subsection{Stratospheric methane abundance \label{sec:heat-sources-methane}}

The abundance of methane is rather poorly constrained on both planets. We explore the impact of the methane abundance on our simulated profile to assess if the mismatch between models and observations could be solved by setting a specific methane abundance within current observational errors.

\begin{figure}[h!]
        \centering
        \includegraphics[width=.40\textheight]{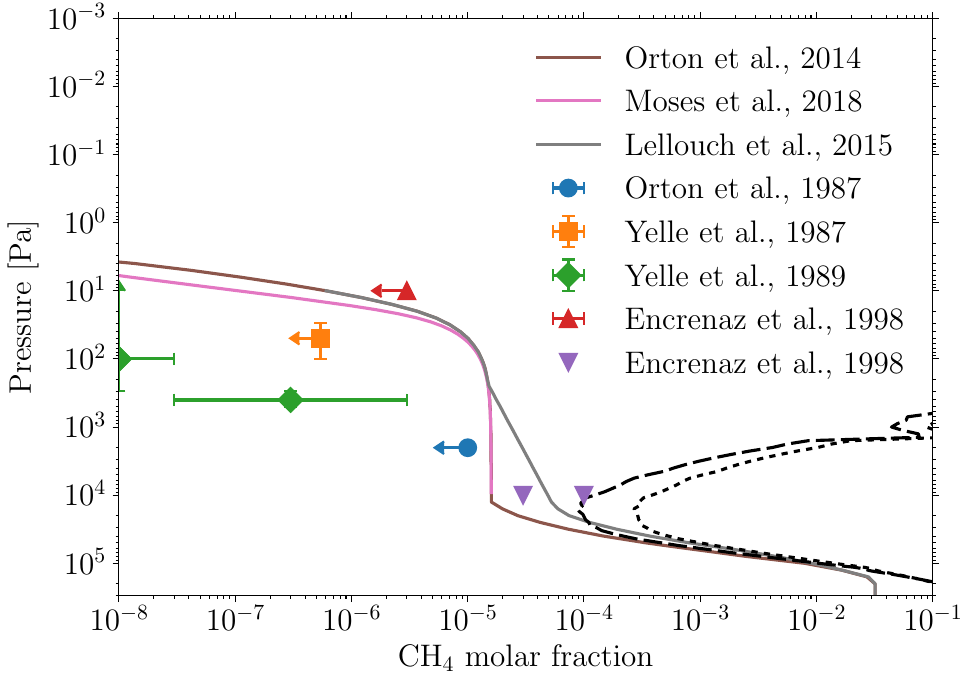}
        \caption{\label{fig:methane_uranus}Methane volume mixing ratio in the atmosphere of Uranus estimated by different observations and models. The dotted line represents the liquid saturated CH$_4$ vmr calculated from \citet{Lindal1987} temperature profile and the dashed line is the ice saturated CH$_4$ vmr. The nominal CH$_4$ vmr profile used in our model is \citet{Lellouch2015} below the 100 Pa level and \citet{Moses2018} above the 100 Pa level.}
\end{figure}

Various estimates of the methane abundance obtained in Uranus' atmosphere are summarised in fig.\ref{fig:methane_uranus}. The general observed trend is a strong decrease above the tropopause level, where methane decreases from typically $\sim$10$^{-4}$ at 100~hPa to below 10$^{-6}$ at the 1~hPa level. This trend is qualitatively reproduced by the seasonal photochemical model of \citet{Moses2018,Moses2020}, although that model significantly exceeds the methane abundance derived from UVS/Voyager~2 by \citet{Yelle1987, Yelle1989} at 1--10~hPa. Assuming a lower value of the eddy diffusion coefficient in the photochemical model (hence resulting in a lower homopause level) can match the UVS methane data. However, this results in a strong underestimation of the  acetylene abundance derived from UV/visible spectrum and thermal infrared (see discussion in \cite{Moses2005} and references therein). This suggests that the homopause level may vary with latitude and/or season, or that the methane abundances derived from UVS were strongly underestimated.
Measurement discrepancies in the methane abundance are also found in the lower stratosphere, where \citet{Lellouch2015} derived a volume mixing ratio (vmr) equal to $9.2 \times 10^{-5}$ near 100 hPa from Herschel far infrared and sub-mm observations (compatible with analyses of ISO measurements by \citet{Encrenaz1998}), while the analysis of  Spitzer/IRS observations by \citet{Orton2014} is consistent with a much smaller abundance, of $1.6 \times 10^{-5}$, at that pressure level. 


\begin{figure}[h]
        \centering
        \includegraphics[width=.40\textheight]{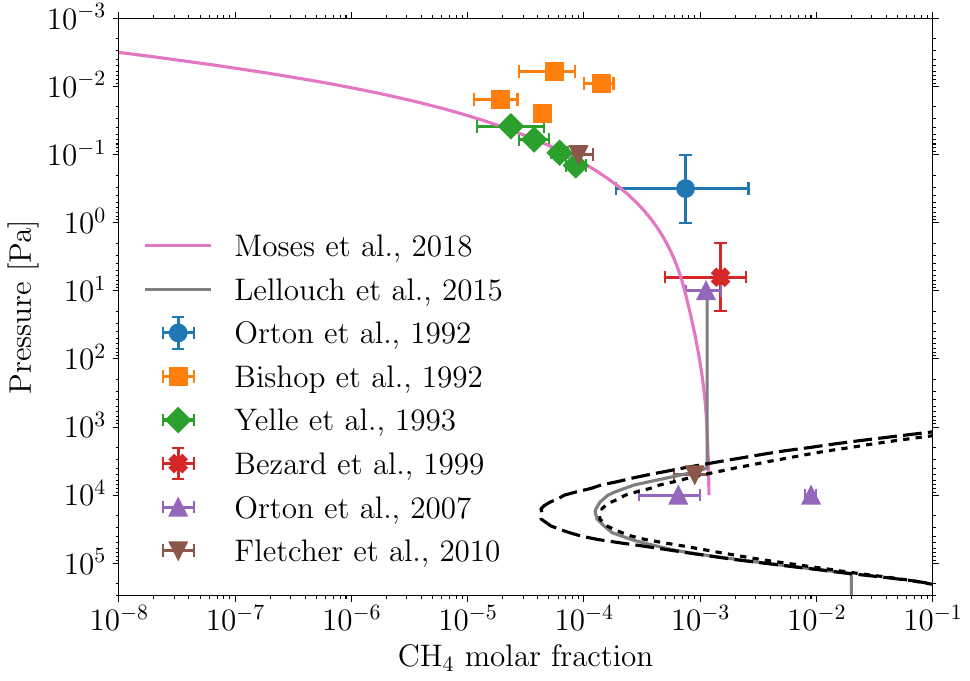}
        \caption{\label{fig:methane_neptune}Methane volume mixing ratio in the atmosphere of Neptune retrieved by different observations and models. The dotted line represents the liquid saturated CH$_4$ vmr calculated from \citet{Lindal1992} temperature profile and the dashed line is the ice saturated CH$_4$ vmr. The CH$_4$ vmr profile used in our model is \citet{Lellouch2015} between 1500 and 100 hPa from and \citet{Moses2018} above 100 hPa level. Below the 1500 hPa, the CH$_4$ vmr value from \citet{Irwin2019b} is assumed.}
\end{figure}

For Neptune, several estimations of the stratospheric CH$_4$ vmr profile exist (fig.\ref{fig:methane_neptune}). From ground-based spectroscopic infrared observations, CH$_4$ vmr estimations below the 1 Pa level and above the tropopause seem to be in agreement with each other with a value of $\sim1 \times 10^{-3}$. At lower pressures, a discrepancy between observations appears. From Voyager~2/UVS solar occultation lightcurves, \citet{Bishop1992} constrained the abundance of methane to be in the range $0.2-1.5 \times 10^{-4}$ between 0.006 and 0.025 Pa. However, with the same dataset, \citet{Yelle1993} derived almost the same values but at higher pressures ($\sim$0.1 Pa). Using the disk-averaged infrared spectra obtained by Akari, \citet{Fletcher2010} derived a similar value at this pressure level. The photochemical model from \citet{Moses2018} is also consistent with the latter estimations of CH$_4$ vmr at this level and also reproduces the CH$_4$ vmr estimated in the low stratosphere. 

The nominal methane profile used in our model consists in the \citet{Lellouch2015} profile below the 100 Pa level for Uranus (1000 Pa for Neptune) and \citet{Moses2018} above it. On Neptune, the methane tropospheric value is set to 4\% following \citet{Irwin2019b}). However, the aforementioned discrepancies in different observed methane values in the lower stratosphere and/or in the homopause altitude level (fig. \ref{fig:methane_uranus} \& \ref{fig:methane_neptune}) leave room to test other methane profiles and evaluate their influence on the temperature profile. We also test unrealistically high methane abundance profiles to evaluate what would be the amount of methane needed to match the observed temperatures. For Uranus, tests were carried out by multiplying by 2, 4, 6, 8 and 10 our nominal case of CH$_4$ vmr above the methane condensation level, with the constraint of never exceeding CH$_4$ local liquid saturation. The highest methane abundance tested for Uranus is similar to the mean CH$_4$ vmr retrieved on Neptune's lower stratosphere. In the case of Neptune, the methane homopause level is already above the top of the model ($\sim$5 Pa). So, only tests by multiplying the CH$_4$ vmr above the a priori methane condensation level have been done without exceeding CH$_4$ local liquid saturation on Neptune (fig. \ref{fig:methane_neptune}). 

\begin{figure}[h]
    \centering
    \includegraphics[width=.45\textheight]{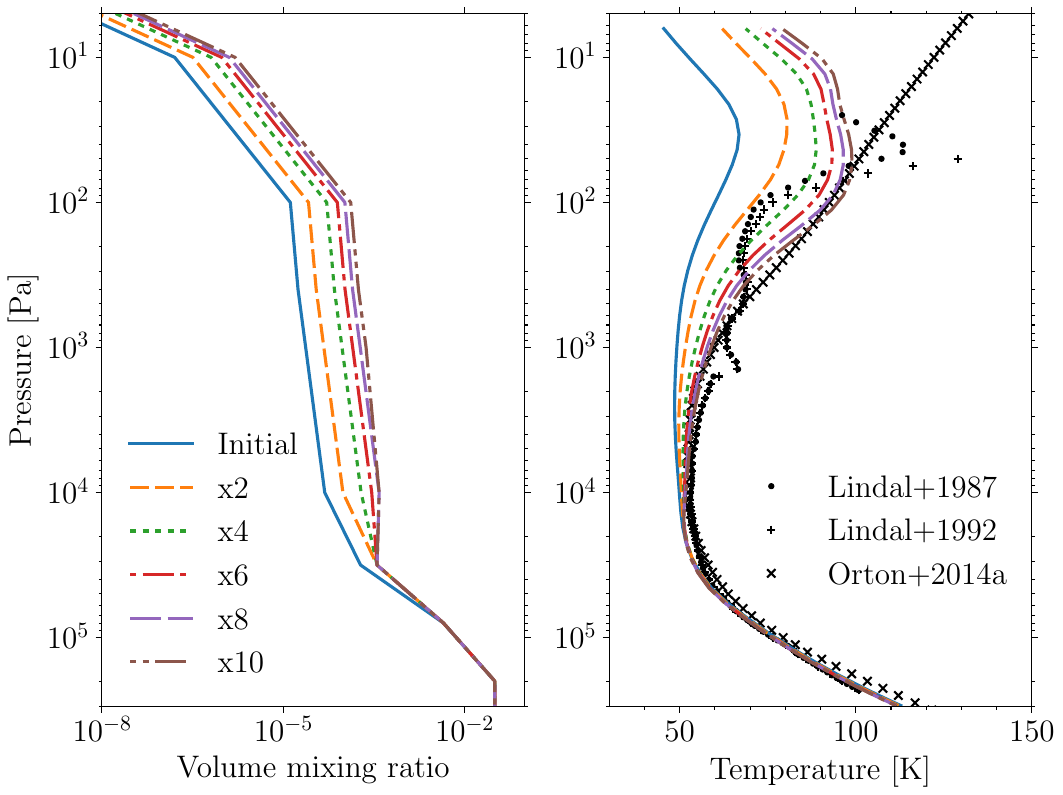}
    \caption{\label{fig:methane_heating_uranus}Left: CH$_4$ vmr profile on Uranus (solid blue line, see section \ref{sec:initial-profile}), with 2 (short dashed orange curve), 4 (dotted green curve), 6 (dotted-dashed red curve), 8 (long dashed violet curve) and 10 (double dotted-dashed maroon curve) times the nominal abundance. Right: the corresponding simulated temperature profiles. The black symbols are the observed temperatures described in fig.\ref{fig:noaerosol} .}
\end{figure}

The tests presented in this section have been made without any aerosol/cloud layers. Results obtained with different values of methane abundance show that the stratosphere of Uranus is highly sensitive to the amount of methane (fig.\ref{fig:methane_heating_uranus}). A value of $10^{-4}$ for the CH$_4$ vmr throughout the lower stratosphere, which corresponds to a saturated case at the tropopause, can sufficiently warm these levels to match the observations. This amount is too large compared to the globally-averaged observations \citep{Lellouch2015} (3 to 10 times higher depending on the pressure level). One could argue the possibility of a greater local methane abundance at the location of the Voyager~2 radio-occultation profile. However, the globally-averaged temperature being similar to the Voyager~2 temperature, this assertion is unlikely but strong and temporal methane gradients as for C$_2$H$_2$ \citep{Roman2022} can exist. Supplementary tests have been made by increasing only the homopause of the nominal profile from 100 to 10 Pa by maintaining the vertical gradient between 100 and 200 Pa. A warming has been observed but it remained confined to the last levels of our model (for pressures lower than $\sim$50 Pa). The Bond albedo obtained for all abundances tested is similar to the clear-sky Bond albedo (see section \ref{sec:initial-profile}).

\begin{figure}[h]
    \centering
    \includegraphics[width=.45\textheight]{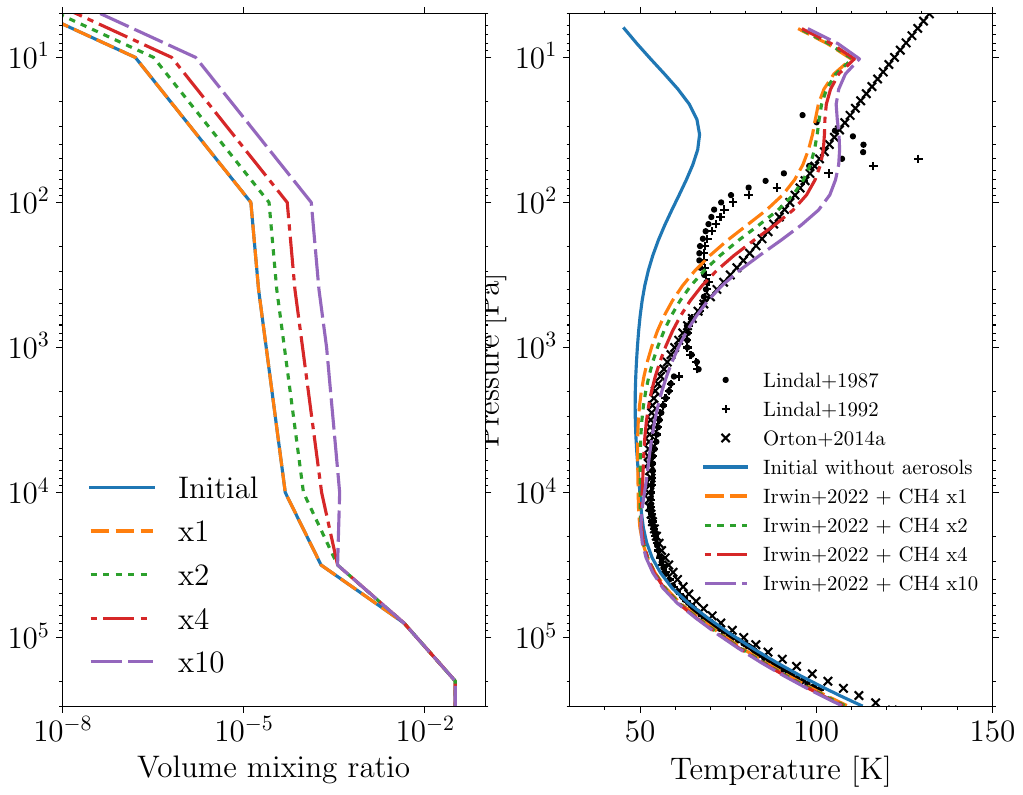}
    \caption{\label{fig:aerosols_methane_uranus} Left: CH$_4$ vmr profile on Uranus (blue line and short dashed orange line) or with that abundance multiplied by a factor of 2 (dotted green line), 4 (dotted-dashed red line), 10 (long dashed violet line). Right: the corresponding simulated temperature profiles with the \citet{Irwin2022} haze scenario. The solid blue line corresponds to the initial case without hazes (see section \ref{sec:initial-profile}) with a nominal CH$_4$ vmr profile. The black symbols are the observed temperatures described in fig.\ref{fig:noaerosol}.}
\end{figure}

By adding the best-fit cloud/aerosol scenario, the amount of additional methane necessary to warm the stratosphere is much less important, or even not needed considering that the temperature simulated only with the cloud model is already close to the Voyager~2 profiles. An amount 2 to 4 times higher than our reference profile, still consistent with the range of observations (fig.\ref{fig:aerosols_methane_uranus}), enables to match the temperature profile of Voyager~2. Thus, a higher abundance of methane can partly account for the warm observed temperatures on Uranus.

In any of the tested scenarios described above, the combination of cloud/haze scenarios and various methane abundance profiles remain insufficient to reproduce the observed temperature on Uranus at pressures lower than 30~Pa. An additional heating is required at the top of the model to reduce the gap between the simulated temperature and the observed one.

\begin{figure}[h]
    \centering
    \includegraphics[width=.45\textheight]{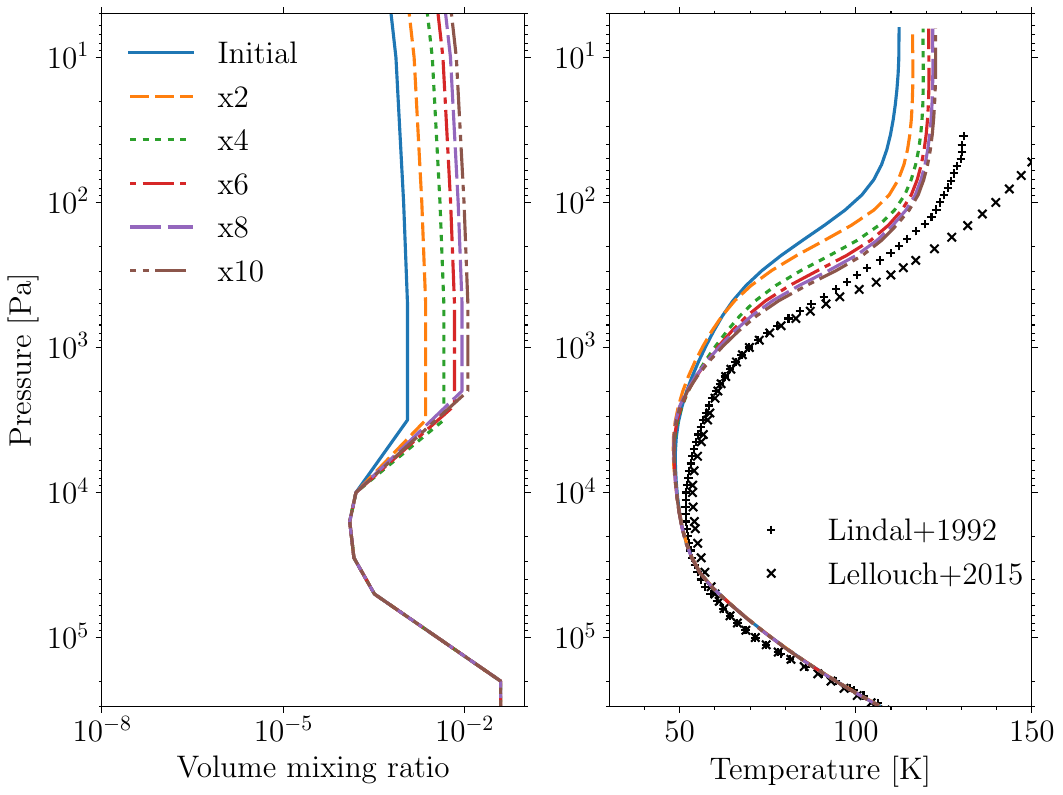}
    \caption{\label{fig:methane_heating_neptune}Left: CH$_4$ vmr profile on Neptune with the nominal CH$_4$ vmr profile (solid blue line, see section \ref{sec:initial-profile}), or with that abundance multiplied 2 (short dashed orange curve), 4 (dotted green curve), 6 (dotted-dashed red curve), 8 (long dashed violet curve) and 10 (double dotted-dashed maroon curve) above the methane condensation level. Right: the corresponding simulated temperature profiles. The black symbols correspond to the observed temperatures described in fig.\ref{fig:noaerosol} .}
\end{figure}

On Neptune, the atmosphere is far less sensitive to the addition of higher methane amounts (fig.\ref{fig:methane_heating_neptune}) at any level because the spectral windows of the methane are already saturated. To be close to the observed temperature, the methane abundance would have to be 100 times greater than that measured. Thus, a higher methane abundance cannot be the solution to the "energy crisis" on Neptune, even with additional heating from aerosol.

\subsection{Thermospheric conduction \label{sec:heat-sources-conduction}}

The thermospheres of Uranus and Neptune are well-known for the "energy crisis" problem. The temperature expected from solar heating is inconsistent with the observed ones at these levels \citep{Melin2019}. A temperature of 750 K has been measured at the expected level of the thermopause by UV stellar and solar occultations on Uranus \citep{Broadfoot1986} and UV solar occultation on Neptune \citep{Broadfoot1989} during the flyby of Voyager~2. Since then, a significant and constant decrease in temperature, based on several measurements of H$_3^{+}$ from ground-based observations, has been observed on Uranus \citep{Melin2019}. Since 1992, the cooling rate was 8K per terrestrial year such that in 2018, the temperature reached 486 K. In the case of Neptune, no measurement of the thermospheric temperature has been made since the Voyager~2 flyby. However, \citet{Moore2020} suggest from H$_3^{+}$ upper limits and models that the upper atmosphere may have significantly cooled since the Voyager~2 era. In any case, the thermosphere is warmer than expected. Atomic and molecular hydrogen being inefficient radiators, the heat energy on the thermosphere is lost by radiative cooling of H$_3^{+}$ and is carried downward by conduction to lower levels. Our model does not cover the thermosphere. However, the stratosphere of both planets is contiguous to the warm thermosphere and thus their upper part can be warmed by conduction. 

The heat flux $Q_c$ resulting from the thermospheric conduction is approximated as:

\begin{equation}\label{eq:flux_conduction}
    Q_c = -k \frac{\Delta T}{\Delta z}
\end{equation}

where $k$ is the thermal conduction coefficient of the atmosphere, $\Delta T$ is the temperature range over a thickness $\Delta z$ located between the bottom and the top of the thermosphere. The thermal conduction coefficient $k$ is calculated by the semi-empirical formula from \citet{Mason1958} for a gas mixture of orthohydrogen, parahydrogen at equilibrium and helium using data from \citet{Mehl2010} and \citet{Hurly2007}. From these results, an interpolation formula expressed as $k = AT^S$ is used to determine the conduction flux from the thermosphere to add at the top of our model. Between 20 and 200 K, the coefficient $A$ is equal to 1.064$\times$10$^{-3}$ J.m$^{-1}$.s$^{-1}$.K$^{-(S+1)}$ and $S$ to 0.906. 

To determine the $Q_c$ flux at the bottom of the thermosphere, we need to estimate the temperature gradient in the thermosphere, above our model top. In the case of Uranus, two temperature gradients have been tested on the basis of the available observations (table \ref{tab:intheat}). 
The first one (UC1) corresponds to the values deduced from stellar and solar occultations by \cite{Broadfoot1986} (reworked by \citet{Melin2019}). The second gradient (UC2) is that obtained by \citet{Herbert1987}, also derived from stellar and solar occultations during the Voyager~2 flyby. In the first case, the conductive flux obtained is equal to 5.90$\times$10$^{-8}$ W.m$^{-2}$.km$^{-1}$ whereas in the second case, the conductive flux is much higher due to a greater temperature gradient (1.00$\times$10$^{-6}$ W.m$^{-2}$.km$^{-1}$). Unrealistic cases were tested in order to assess the conductive flux required to heat the stratosphere and bring the observed temperatures and the simulated ones into agreement. On Uranus, a conductive flux of 1.36$\times$10$^{-3}$ W.m$^{-2}$, corresponding to a thermospheric gradient of +7 K.km$^{-1}$, is required, and in the case of Neptune, the flux must reach 1.56$\times$10$^{-2}$ W.m$^{-2}$, corresponding to a temperature gradient of +60 K.km$^{-1}$.

For Neptune, the influence of thermospheric conduction is much less plausible according to observations.
Indeed, for the pressures considered above our model top, between 10 and 0.04 Pa \citep{Yelle1993}, there is a vertically thick isothermal temperature zone (stratopause) but poorly constrained. It is only for pressures below 0.04 Pa that we enter the thermosphere with a positive temperature gradient. If we disregard this isothermal zone, the conduction flux resulting from this gradient is equal to 7.38$\times$10$^{-8}$ W.m$^{-2}$.km$^{-1}$ (model NC1 in table \ref{tab:intheat}).

\begin{table}[ht]
  \centering
  \begin{tabular}{ l | c | c | c | }
     Name & $\Delta T$ (K) & $\Delta z$ (km) & $ Q_{c,bottom}$ (W.m$^{-2}$) \\ \hline
     UC1 & 170-500 & 1100  & 6.5$\times$10$^{-5}$ \\ \hline
     UC2 & 150-500 & 260 & 2.62$\times$10$^{-4}$ \\ \hline
     UC3 & 150-500 & 50 & 1.36$\times$10$^{-3}$ \\ \hline
     NC1 & 150-750 & 1450 & 1.07$\times$10$^{-4}$ \\ \hline
     NC2 & 150-750 & 10 & 1.56$\times$10$^{-2}$ \\
  \end{tabular}
  \caption{\label{tab:intheat} Heat fluxes tested for Uranus and Neptune as a function of the temperature range in the thermosphere and the thickness of this layer. For Uranus, two values of conductivity were tested (see text). $Q_{c,top}$ corresponds to the value of conduction flux at the top of the thermosphere and is equal to 0. 
  }
\end{table}

From these heat fluxes, the heating rate from thermospheric conduction is parameterised as follows:

\begin{equation}
    \frac{\partial T}{\partial t} = \frac{1}{\rho c_p} \frac{\partial}{\partial z} \left ( k \frac{\partial T}{\partial z} \right)
\end{equation}
with $c_p$ the specific heat capacity at constant pressure, $\rho$ the density. The thermospheric conduction flux is fixed at the last level of our model where the formula can be rewritten as:

\begin{equation}
    \frac{\Delta T}{\Delta t} = \frac{1}{\rho c_p} \frac{Q_{c,bottom}}{\Delta z}
\end{equation}

\begin{figure}[h]
    \centering
    \includegraphics[width=.45\textheight]{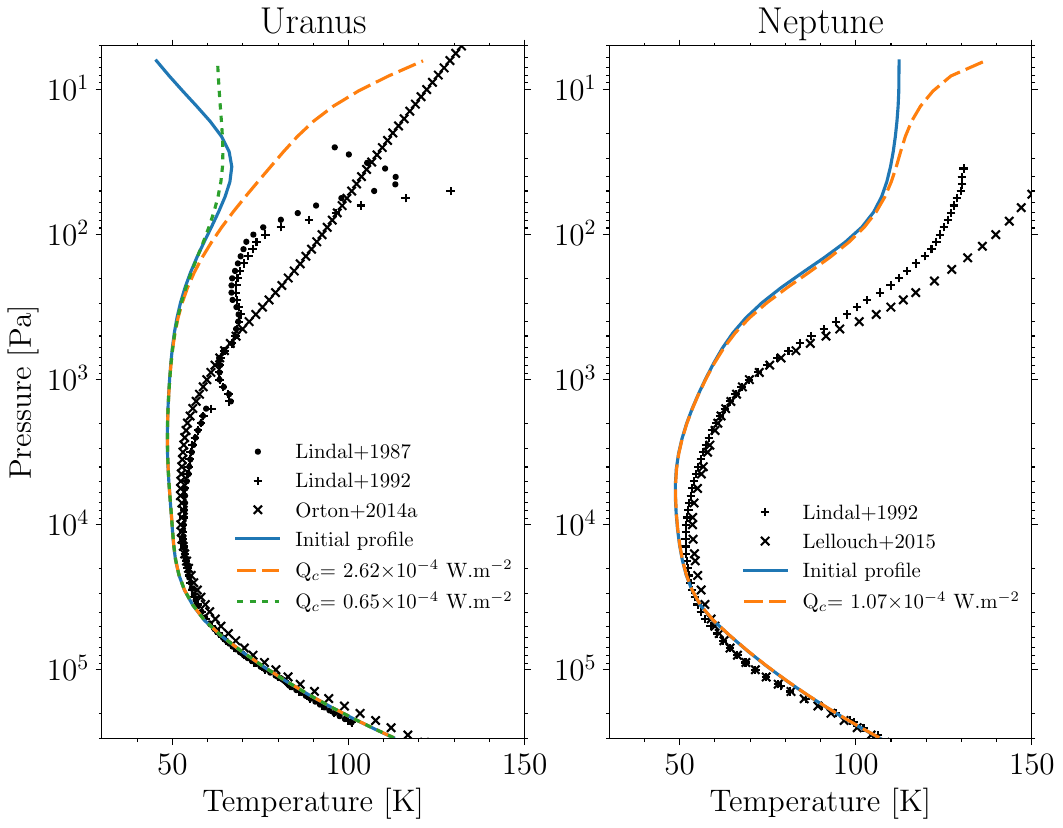}
    \caption{\label{fig:conduction}Scenario with no hazes including heat conduction. Left: simulated temperature profiles on Uranus for the UC1 (dashed orange curve) and UC2 (dotted green curve) thermospheric conduction scenarios. Right: simulated temperature profiles on Neptune for NC1 (dashed orange curve) thermospheric conduction scenario. The solid blue line is the initial case without conduction (section \ref{sec:initial-profile}). The black symbols are the observed temperatures described in fig.\ref{fig:noaerosol}.}
\end{figure}

\begin{figure}[h]
    \centering
    \includegraphics[width=.45\textheight]{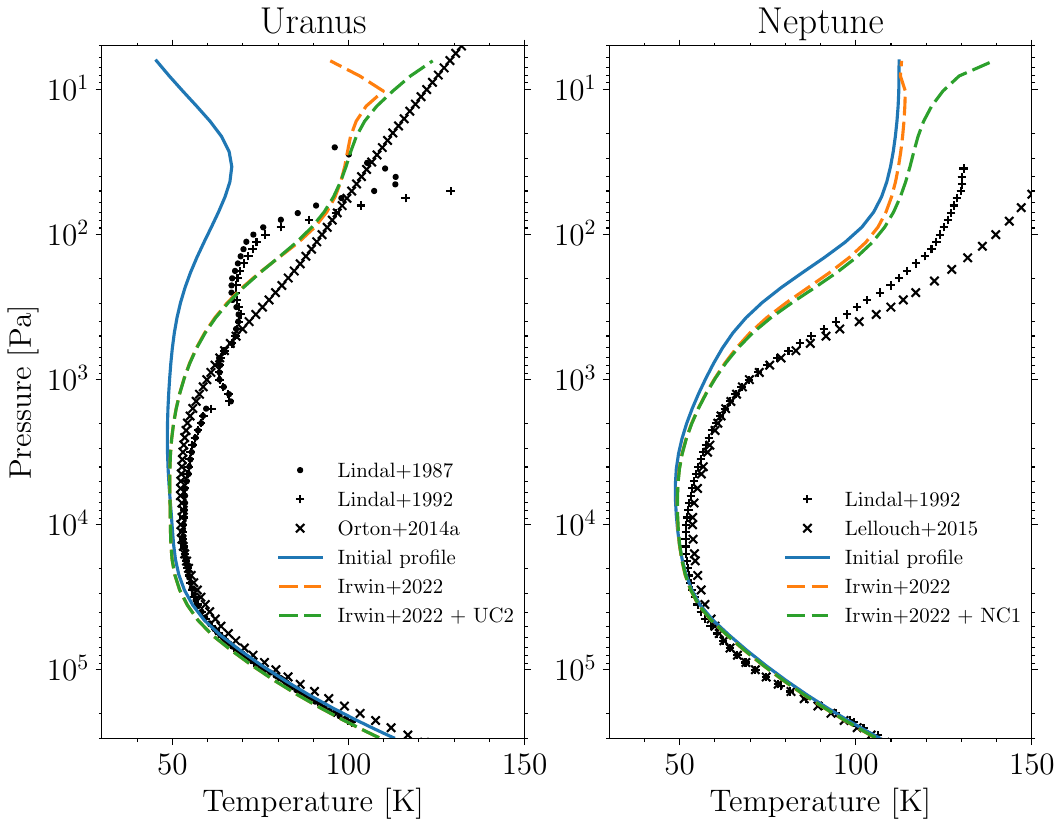}
    \caption{\label{fig:aerosols_conduction}Left: simulated temperature profiles on Uranus with the \citet{Irwin2022} haze scenario but no conduction (orange long-dashed line) and with haze and thermospheric conduction with the UC2 flux scenario (green dashed line). Right: same for Neptune (but with the NC1 flux scenario for the green dashed line). The solid blue line is the initial temperature (section \ref{sec:initial-profile}) and the black symbols are the observed temperatures described in fig.\ref{fig:noaerosol} .}
\end{figure}

On Uranus, in the case without aerosols (fig.\ref{fig:conduction}), the top of our model (at pressures below 50 Pa) is sensitive to a conductive heat flux. All cases show that the thermospheric conduction can not warm levels below the 50 Pa level on Uranus and Neptune. This is expected, as the vertical temperature gradient becomes very small and as radiative effects start to dominate heat exchanges at these levels. In the UC2 case, the temperature gain obtained from heat conduction on the last pressure level (5 Pa) of our model is 70 K while it is of the order of 10K in the UC1 case. With aerosols (fig.\ref{fig:aerosols_conduction}), the UC2 scenario allows the temperature to be increased by 15 K at the top of the model and to keep a more realistic temperature gradient in the upper stratosphere. Concerning Neptune, the atmosphere remains insensitive to such a flux except on the top of our model (for pressures below 10 Pa). Moreover, the methane already dominates radiative exchanges at these pressure levels (unlike Uranus, where the methane homopause resides at much higher pressures). The same is true by adding the aerosol scenario. In any case, thermal conduction can not be a solution to the "energy crisis" in the middle stratosphere of Neptune. In addition, the thermospheric conduction creates a temperature gradient in the upper stratosphere inconsistent with the isothermal zone present above the 10 Pa level \citep{Yelle1993}.

\subsection{Discussion \label{sec:heat-sources-discussion}}

Previous radiative-convective models \citep{Appleby1986, Friedson1987, Wang1993, Marley1999, Greathouse2011, Li2018} failed to correctly reproduce the temperature profile observed by Voyager~2 without additional heat sources. They all featured a temperature profile from the tropopause to the stratosphere that was too cold compared to the data, like our initial profile on Uranus and Neptune (fig.\ref{fig:noaerosol}). 

Adding more methane in the stratosphere was one of the first hypotheses considered as the missing heat source for these two planets, especially for Neptune. 
\citet{Marley1999} investigated this issue on Uranus by setting realistic methane values. Due to poor constraints on the methane abundance on Uranus, a wide range of values were tested. Assuming small abundances of 10$^{-7}$ to 10$^{-10}$ (consistent with the Voyager~2 UVS results by \citet{Herbert1987} and \citet{Stevens1993}) in the upper stratosphere (between 3 and 0.3 Pa), they found that methane radiates downwards and warms the atmosphere above 50 Pa by up to 15 K. By using similar values of methane abundance at these levels and without hazes, we reproduce a similar result on Uranus where an increase of 10 K at 10 Pa is seen. For the low stratosphere and the tropopause, a CH$_4$ vmr 10 times higher than observed is necessary to warm these levels in both \citet{Marley1999} and our simulations. In the case of Neptune, 
\citet{Greathouse2011} found that an increase of the CH$_4$ abundance by a factor of 8 was needed to bring the retrieved temperature and the simulated ones into agreement, but in our simulations, an even higher abundance (more than 10 times the observed profile) would be needed (fig.\ref{fig:methane_heating_neptune}). However, as explained here and by the authors, such a high abundance is inconsistent with respect to previous observations \citep{Lellouch2010}.

With the difficulty of reproducing the observed temperature profile with a realistic methane abundance, the hypothesis of aerosols that absorb the solar flux, locally or globally, was quickly proposed \citep{Appleby1986, Lindal1987, Pollack1987, Bezard1990, Rizk1990}. \citet{Appleby1986} analysed the geometric albedo spectrum from the International Ultraviolet Explorer and found sufficient heating by adding an aerosol layer distributed uniformly between 500 and 3$\times$10$^{-2}$ hPa which absorbed 15\% of the total solar irradiance on Uranus. On Neptune, they found that adding an aerosol heating source in the radiative zone does not totally solve the gap between the observed and simulated temperature profiles. However, with the reanalysis by \citet{Karkoschka1994} of the IUE spectrum, \citet{Marley1999} found that the reanalysed spectrum was inconsistent with the aerosol heating from \citet{Appleby1986}.

\citet{Marley1999} used the haze scenario from \citet{Rages1991} to determine the influence of hazes on the temperature profile of Uranus. They found that the simulated temperature was similar to their 'cold baseline profile' (a case without hazes). 
This lack of heating by haze in their model may be explained by the assumption of small stratospheric haze particles in the scenario of \citet{Rages1991} (decreasing from a radius of 0.05 $\mu$m at 500 Pa to 0.01 $\mu$m at 40 Pa). This drop in particle size has the effect of favouring scattering rather than absorption at lower pressures. By increasing the imaginary index of refraction of \citet{Rages1991} haze scenario, the aerosols become sufficiently darker in the ultraviolet to warm the stratosphere \citep{Marley1999}. However, the observed UV geometric albedo is not reproduced \citep{Marley1999} and the analysis of a Raman scattered Lyman-$\alpha$ emission in ultraviolet \citep{Yelle1987} shows that the atmosphere above the 0.5 hPa level seems to be clear on Uranus. \citet{Marley1999} thus rejected the hazes as a possible solution of the "energy crisis" on Uranus. The haze scenario from \citet{Irwin2022}, which is derived from the analyses of reflected spectra between 0.4 and 2.3 $\mu$m, can warm the stratosphere of Uranus and be close to the Voyager~2 profile (fig.\ref{fig:irwin2022}) with a particle radius sufficiently large to lead absorption and do not need to be dark in the UV. The low amount of methane in Uranus' stratosphere allows any hazes slightly opaque in the visible and NIR light to be an important heat source, which is not the case on Neptune due to higher amounts of methane. 
The nature of these hazes is uncertain but could be similar to tholins or ice tholins \citep{Irwin2022}. Using optical constants of tholins instead of those of \citet{Irwin2022} allows obtaining a similar heating to that produced by Irwin's hazes (fig.\ref{fig:tholins}). The Bond albedo obtained with the tholins (0.29 for Uranus and 0.27 for Neptune) is also close to the observed values. The Bond albedo calculated with the ice tholins is still consistent with the observations for Neptune (0.33) but for Uranus, the ice tholins reflect too much light (0.38). 
The real vertical distribution of the stratospheric hazes is also unknown. A more complex vertical structure is expected with local minima and maxima of optical depth and different particle radius \citep{Toledo2019,Toledo2020}. Then, the efficiency of the heating by the aerosols can be different between the layers compared with our simulations. The aerosols may also be organised into bands in the same way as we see for Jupiter and Saturn. On Uranus, several bands have been revealed in near-infrared images \citep{Sromovsky2015}.

On the other hand, we note that \citet{Li2018} raised a problem concerning the presence of oxygen-bearing species like carbon monoxide \citep{Cavalie2014}, which can cool the atmosphere and compensate for other heating sources. This hypothesis was not tested here and its importance has not been determined. Furthermore, in this study, it was hypothesised that the aerosols were Mie spheres. In the stratosphere of Jupiter, \citet{Zhang2015} showed that fractal aggregate particles produced by coagulation processes dominate the heating at middle and high latitudes contrary to simple Mie aerosol layers. The heating efficiency is better in the case of fractal aggregate hazes because they are optically thinner in the mid-infrared wavelengths and thicker in the visible/ultraviolet \citep{Wolf2010}. A similar configuration of aerosols on ice giants has not been explored yet and would need to be investigated in future studies.

\citet{Rizk1990} proposed that dust from the rings falls near the equator, and behaves optically as black carbon. Using some assumptions on opacity, they found that the 0.1-1 Pa layer can be warmed with a carbon-water ice mix. In our model, by extending this layer from the top to 1000 Pa and by using the optical indices of black carbon \citep{Jager1998}, the dust heating is sufficient to get closer to the temperature observed in the stratosphere not only on Uranus but also on Neptune due its strong absorption continuum in the visible/ultraviolet wavelengths. 
However, this solution suffers from a lack of observation. The concentration, vertical and meridional distribution, particle sizes and their full optical properties remain poorly constrained.
Furthermore, as explained by \citet{Rizk1990}, the ring particles fall only at the equator and can explain the warming only in this region. 
The dust particles would need to be advected from the equator to high latitudes to warm all the stratosphere. 

The effect of thermospheric conduction on stratospheric heating has been little studied to date. For Uranus, \cite{Marley1999} managed to heat the upper stratosphere by 9 K by adding thermospheric conduction (profile B1) to their baseline profile (profile B). 
They note that conduction has an effect on the temperature profile only for pressures below 50 Pa and that its efficiency is sensitive to the abundance of methane in the thermosphere, as in our simulations. Unfortunately, their study does not focus on the values of conductive flux at the top of the model. We can only see that the addition of conduction allows the upper stratosphere to be heated, but in a weak to moderate way, as observed in our simulations. 
The effect of thermospheric conduction on the stratosphere remains uncertain on Uranus due to the uncertainties on the thermospheric temperature gradient. On the one hand, ground-based stellar occultations have shown significant temperature variability in this atmospheric region \citep{Baron1989,Young2001}. On the other hand, measurements during the Voyager~2 flyby are inconsistent by several hundred Kelvins compared with ground-based occultations \citep{Saunders2023}.
In the case of Neptune, \cite{Wang1993} found that conduction had no influence for levels deeper than 10 Pa. We also agree with this finding because methane absorption dominates the heating rates below this level.

To summarise, while the Uranian temperature profile can be reproduced with our 1-D radiative-convective equilibrium model (with a realistic haze scenario), none of the heat sources investigated here can properly warm Neptune's stratosphere. 
The answer to the "energy crisis" on Neptune may lie in its dynamical activity. Important temperature fluctuations have been seen from ground-based stellar occultations in the high stratosphere/low thermosphere of Neptune. \citet{Roques1994} identified these fluctuations as the manifestation of inertia-gravity waves emanating from the lower atmosphere. They show that inertia-gravity waves dissipation can compete with the radiative heating and cooling rates from methane between 3 Pa and 0.03 Pa. On Uranus, these waves could also play a role. We defer the study of the impact of waves to future work.

\section{Seasonal and latitudinal temperature evolution\label{sec:seasonal}}

We describe below the seasonal, 2-D latitude-altitude temperature fields obtained at radiative-equilibrium from simulations including the aerosols layers on Uranus and Neptune from \citet{Irwin2022} keeping fixed with time and the thermospheric conduction (UC2 scenario) for Uranus only (fig.\ref{fig:aerosols_conduction}). Our model assumes methane and other hydrocarbons are horizontally uniform and unchanged over time.

\subsection{Overview of the simulated thermal structure \label{sec:seasonal-simulations}}

\begin{figure}[h!]
\centering
\SetFigLayout{1}{2}
  \subfigure[]{
  \includegraphics[width=.45\textheight]{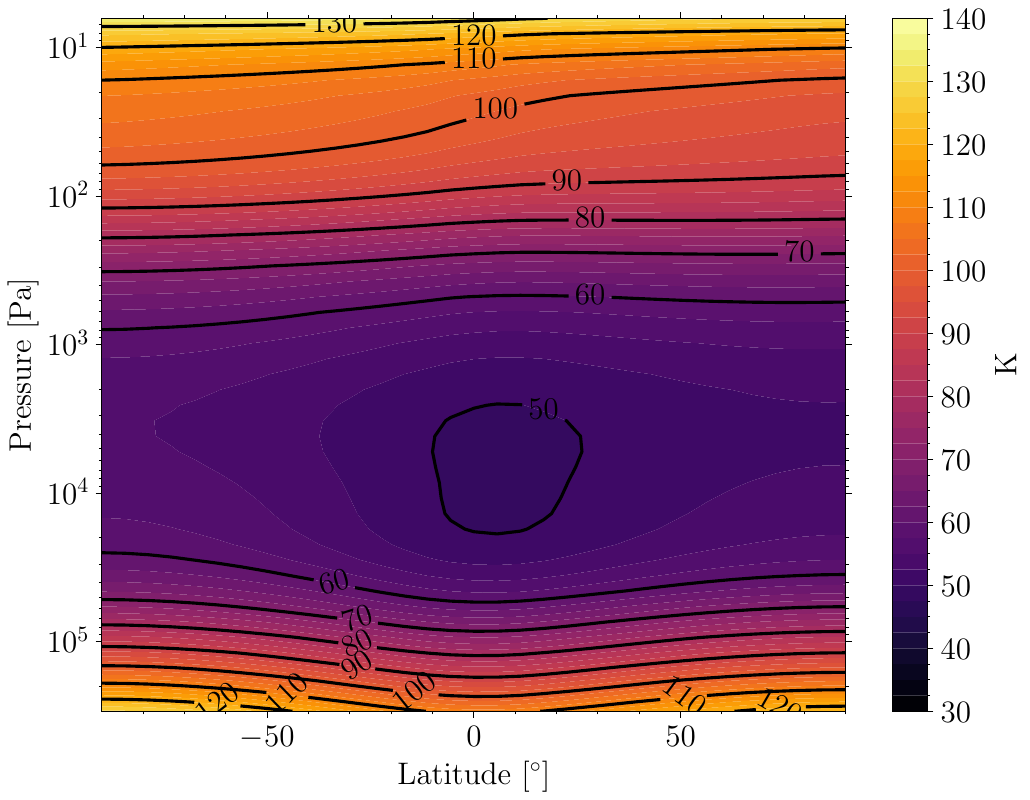}
  \label{fig:uranus-equinox}}
  \hfill
  \subfigure[]{
  \includegraphics[width=.45\textheight]{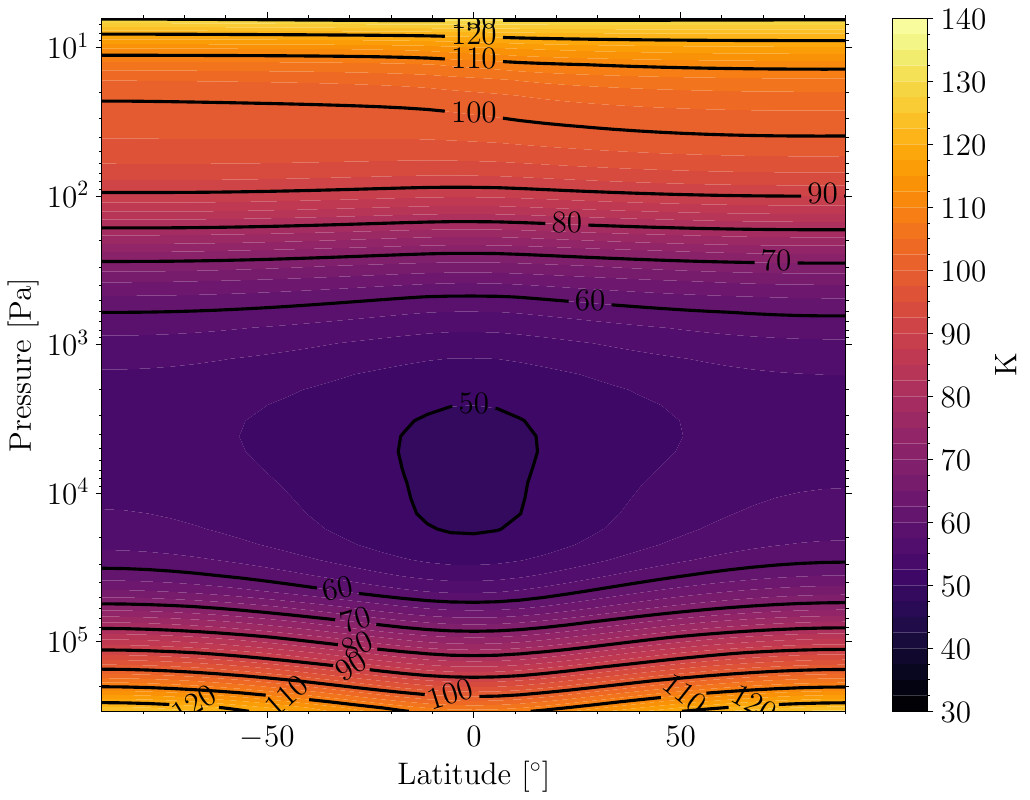}
  \label{fig:uranus-solstice}}
  \hfill
\caption{\label{fig:uranus-nodyn}Uranus: Vertical cross sections of temperature at spring equinox (Ls=0°) (a) and northern summer solstice (Ls=90°) (b). The temperature observed at autumn equinox and winter solstice is almost similar (because of its slightly non circular orbit) as spring equinox and summer solstice but the maximum/minimum temperature is reversed in latitude (fig.\ref{fig:uranus-ls}).}
\end{figure}

\begin{figure}[h]
    \centering
    \includegraphics[width=.45\textheight]{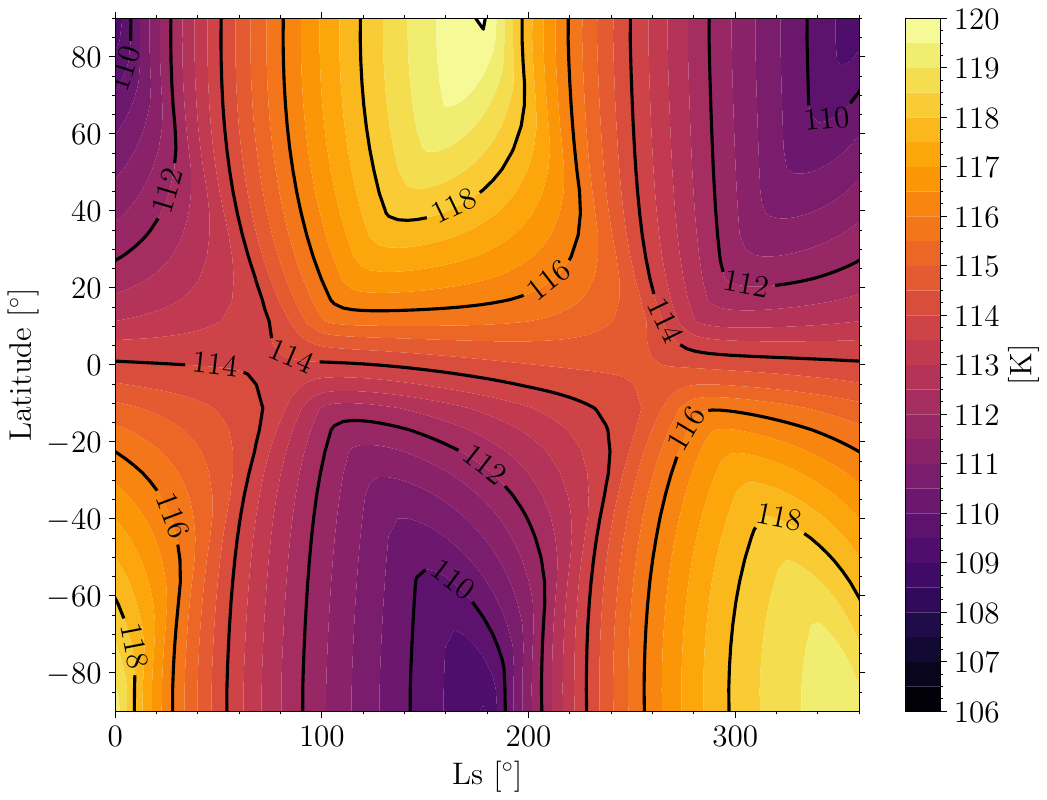}
    \caption{\label{fig:uranus-ls}Evolution of the simulated temperature at 10 Pa on Uranus during one Uranian year.}
\end{figure}

In our simulation of Uranus' stratosphere, maximum latitudinal contrasts (pole to pole) are found to occur at 10~Pa near Ls$\simeq$180° and 0°, during the equinoxes, with a temperature gradient of typically 10 K between northern and southern hemispheres (see fig.\ref{fig:uranus-equinox} and \ref{fig:uranus-ls}). This is shifted by Ls=+90° following the solstices (fig.\ref{fig:uranus-ls}) due to its long radiative time constant \citep{Li2018}. The maximum temperature reached at 10~Pa is 119 K and the minimum is 108 K. A small asymmetry is visible between the northern and southern autumn (resp. spring) where the temperature is $\sim$1.2 K higher (resp. lower) during northern autumn (resp. southern spring). This difference is explained by the eccentricity of Uranus (equal to 0.047) which is high enough to induce a small temperature change. Indeed, the perihelion occurs during the northern autumn equinox (Ls$\simeq$182°) where the seasonal contrast is maximum. The minimum seasonal contrast at 10 Pa occurs at Ls$\simeq$55° and 250°. Below the 1 hPa level, much lower seasonal temperature contrasts are seen. At and below the tropopause level, temperatures are found to be colder at the equator than at the poles throughout the year, consistent with long radiative time scales and with greater insolation at the poles than at the equator on an annual average due to Uranus' extreme obliquity. On average, at 3000 hPa, the temperature meridional gradient reaches $\sim$15 K with a maximum of 123 K at the poles and a minimum of 108 K at the equator. Still, small seasonal changes are observed (on the order of a few kelvins) in the upper troposphere, and the location of the minimum temperature is found to oscillate within $\pm$10° of the equator between the equinoxes. 

\begin{figure}[h!]
    \centering
    \includegraphics[width=.45\textheight]{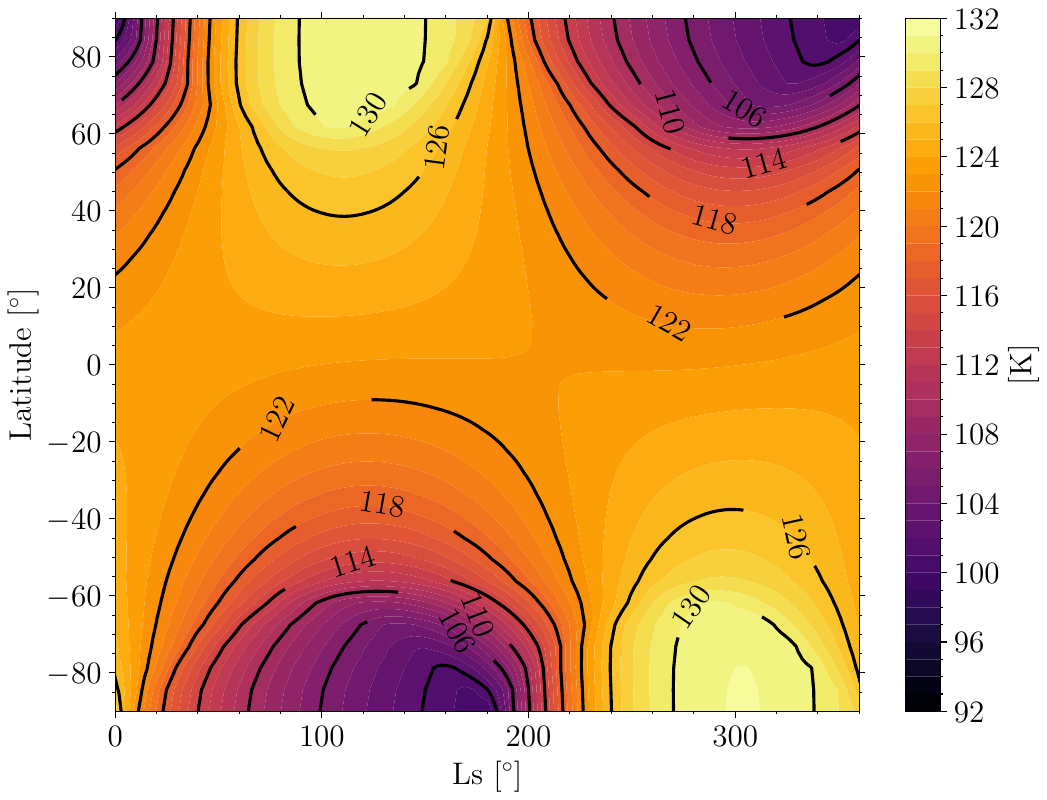}
    \caption{\label{fig:neptune-ls}Evolution of the simulated temperature at 10 Pa on Neptune during one Neptunian year.}
\end{figure}

On Neptune, the maximum seasonal contrast (pole to pole) at 10 Pa is greater than on Uranus due to the higher amount of methane at this level and shorter radiative time scales. The maximum temperature gradient at 10~Pa reaches 27 K between the northern and the southern hemisphere and it occurs at Ls$\simeq$140° and 320°, hence +50° following the solstices (see fig.\ref{fig:neptune-equinox} and \ref{fig:neptune-ls}). This seasonal shift between the solstice and maximum seasonal contrast is similar to that found on Saturn \citep{Guerlet2014, Fletcher2010Saturn}. Contrary to Uranus (and Saturn), Neptune's small eccentricity (equal to 0.009) does not influence its seasonal forcing ($\sim$0.2 K).  The minimum seasonal contrast at 10 Pa occurs near Ls$\simeq$30°/210°. 
The seasonal temperature contrast becomes insignificant (less than 5 K) at pressures greater than 10 hPa. At the tropopause ($\sim$100 hPa), the minimal temperature is found at high latitudes at the end of winter/beginning of spring, reaching 47 K, and the maximum temperature centred at the equator is at 51/52 K. In the troposphere, no significant seasonal variation is seen and the equator-to-pole temperature gradient amounts to $\sim$9 K throughout the year. The maximum temperature at 3000 hPa is 112 K at the equator and the minimum at the poles is equal to 103 K.

\subsection{Comparison to previous radiative-convective models \label{sec:seasonal-previous-models}}

\begin{figure*}[ht!]
\centering
\SetFigLayout{2}{2}
  \subfigure[Uranus: 10 Pa]{
  \includegraphics[width=0.45\textwidth]{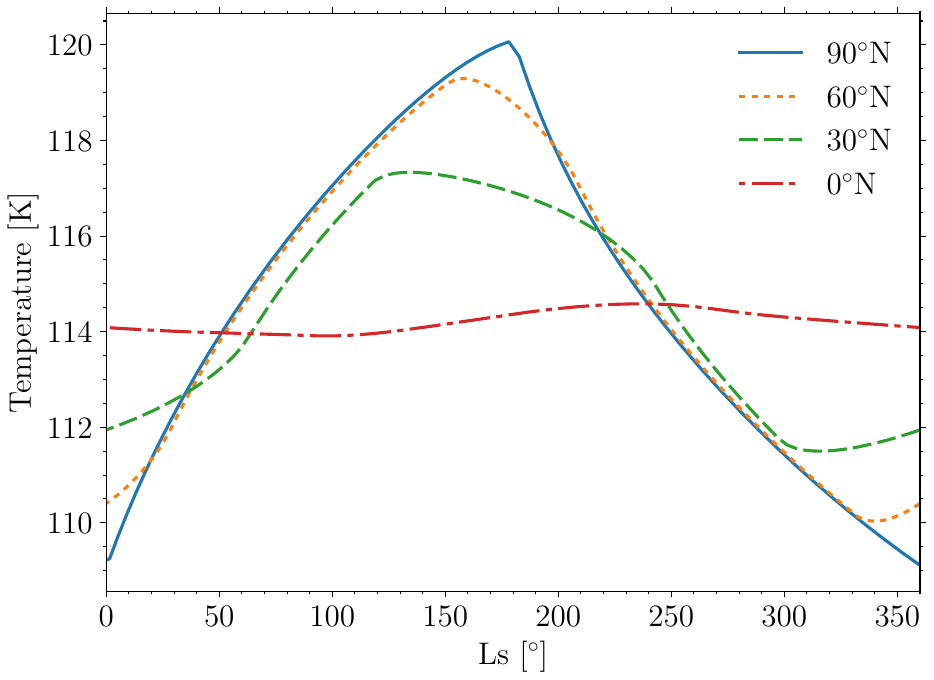}
  \label{fig:uranus-amplitude-10pa}}
  \hfill
  \subfigure[Neptune: 10 Pa]{
  \includegraphics[width=0.45\textwidth]{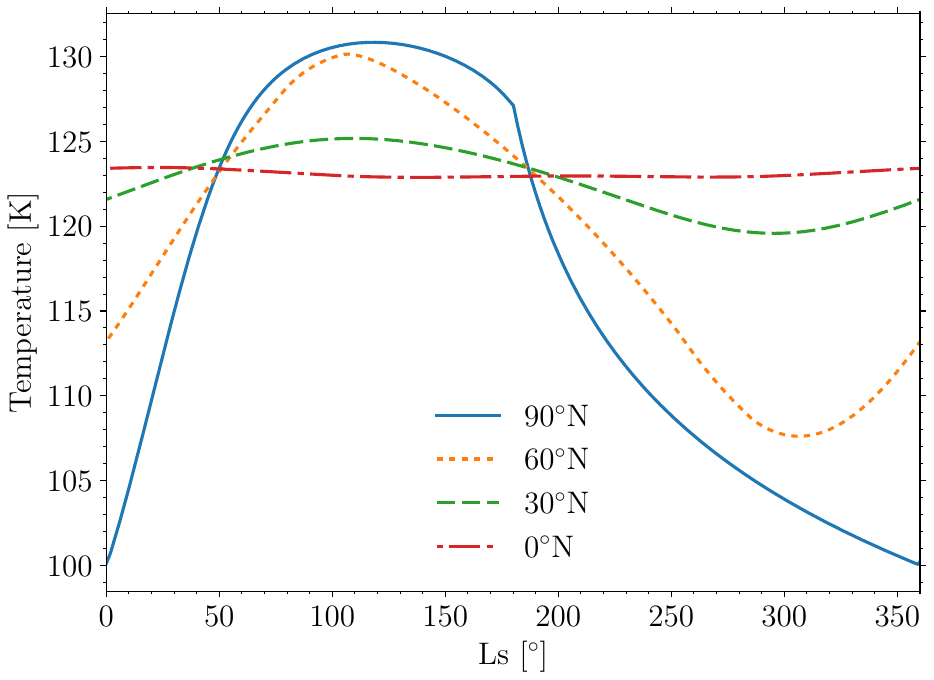}
  \label{fig:neptune-amplitude-10pa}}
  \hfill
  \subfigure[Uranus: 350 hPa]{
  \includegraphics[width=0.45\textwidth]{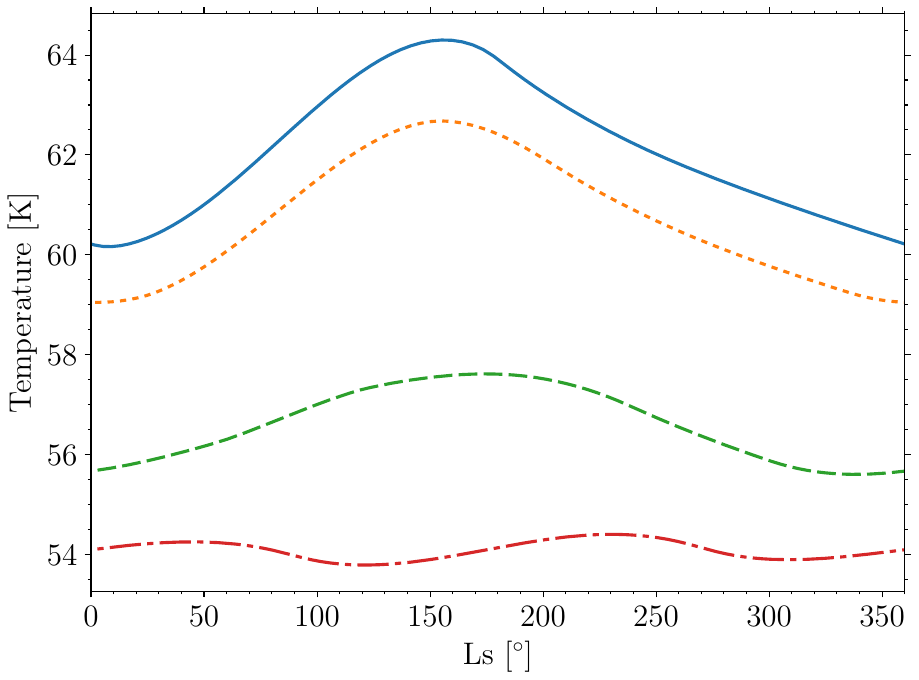}
  \label{fig:uranus-amplitude-300hpa}}
  \hfill
  \subfigure[Neptune: 500 hPa]{
  \includegraphics[width=0.45\textwidth]{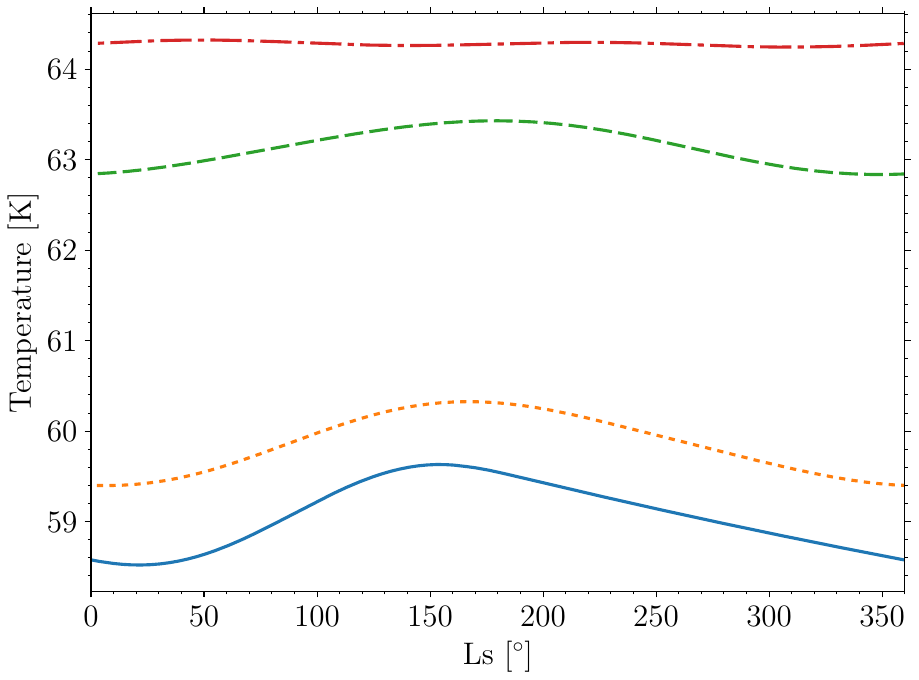}
  \label{fig:neptune-amplitude-500hpa}}
  \hfill
\caption{Evolution of temperature during one planetary year on Uranus (left) at 10 Pa (a) and 350 hPa (c) and on Neptune (right) at 10 Pa (b) and 500 hPa (d) for a latitude of 90°N (blue solid line), 60°N (orange dotted line), 30°N (green dashed line) and 0° (red dashed-dotted line).}
\label{fig:amplitude}
\end{figure*}

Several radiative-convective models were developed during the Voyager~2 era in order to predict the meridional temperature structure on Uranus and Neptune. All previous models produce seasonal temperature variations on Uranus. \citet{Wallace1983} predicted an annual thermal amplitude of the effective temperature (corresponding to $\sim$300 hPa level) on Uranus of 4.9 K at the poles and 0.4 K at the equator (but with an internal heat flux of 0.32 W.m$^{-2}$). This annual contrast at the poles is similar in the simulation of \citet{Bezard1990}, where it amounts to 3.5 K and reaches at most 0.1 K at the equator. In our model, these differences in amplitude between the equator and the poles are also observed (fig.\ref{fig:uranus-amplitude-300hpa}). An annual contrast of 4.1 K at the poles and 0.6 K at the equator is seen and the maximum/minimum temperatures  occur near the equinoxes (at the same level), like on the model of \citet{Wallace1983} and \citet{Bezard1990}. Interestingly, due to the obliquity and the radiative forcing lag, two maxima/minima at the equator are visible (fig.\ref{fig:uranus-amplitude-300hpa}) at pressures higher than 10 hPa. 

\begin{figure}[p]
\centering
\SetFigLayout{1}{3}
  \subfigure[]{
  \includegraphics[width=0.33\textheight]{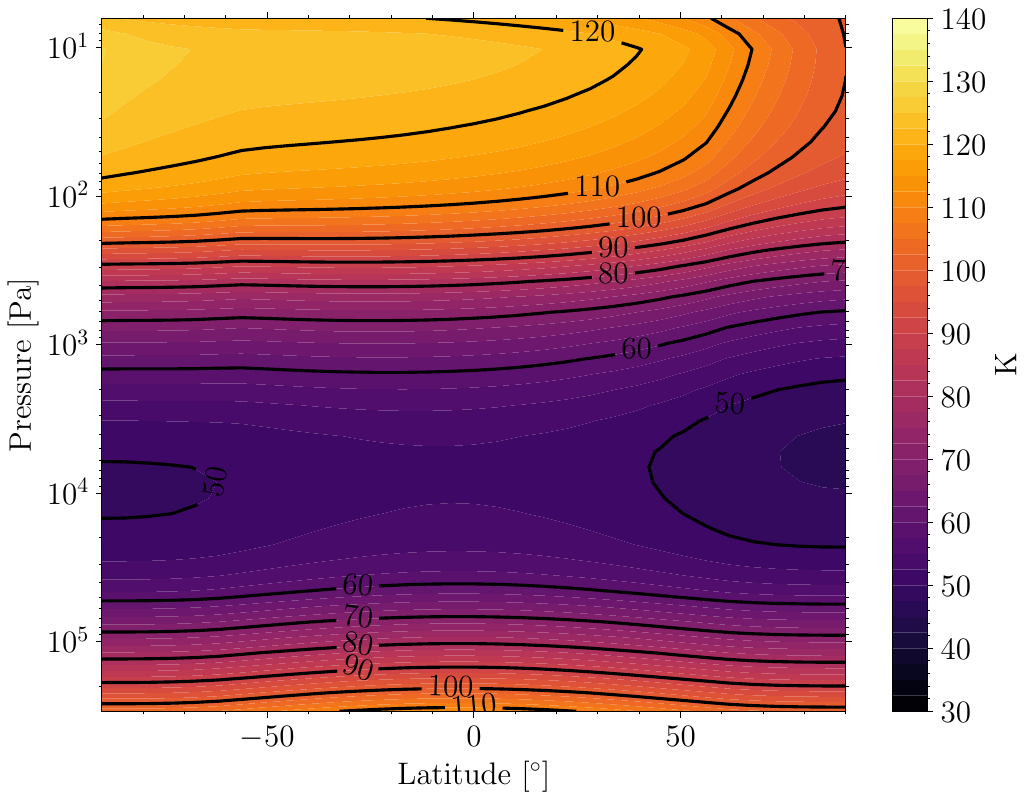}
  \label{fig:neptune-equinox}}
  \hfill
  \subfigure[]{
  \includegraphics[width=0.33\textheight]{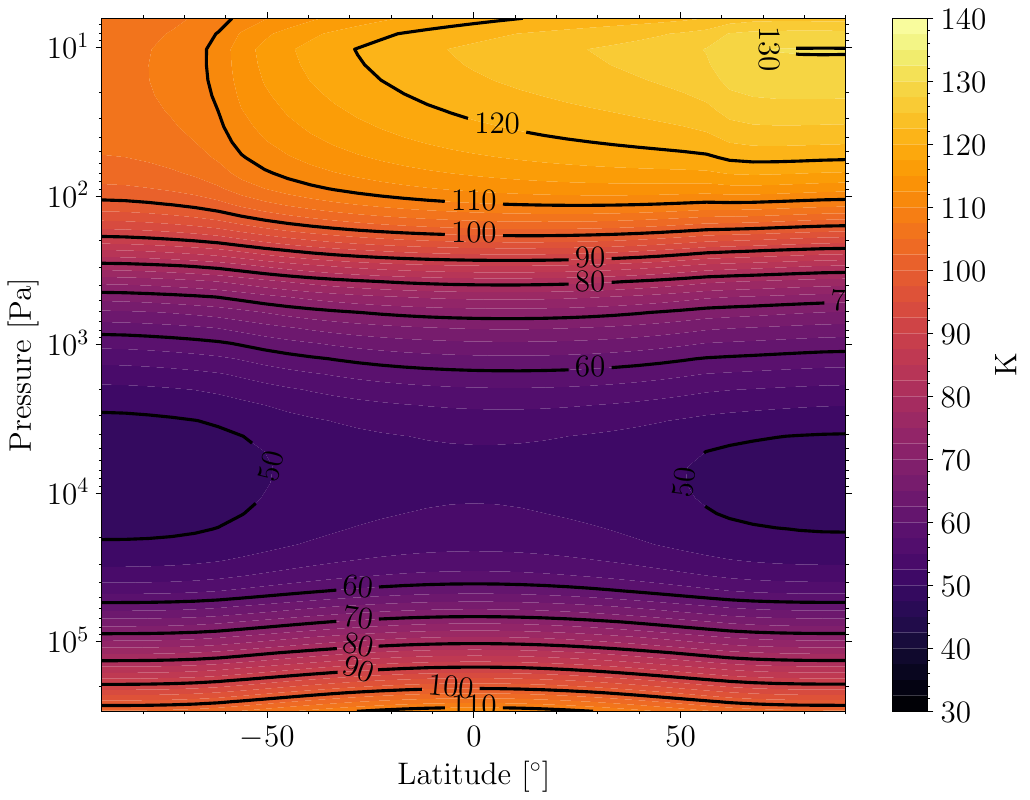}
  \label{fig:neptune-solstice}}
  \hfill
  \subfigure[]{
  \includegraphics[width=0.33\textheight]{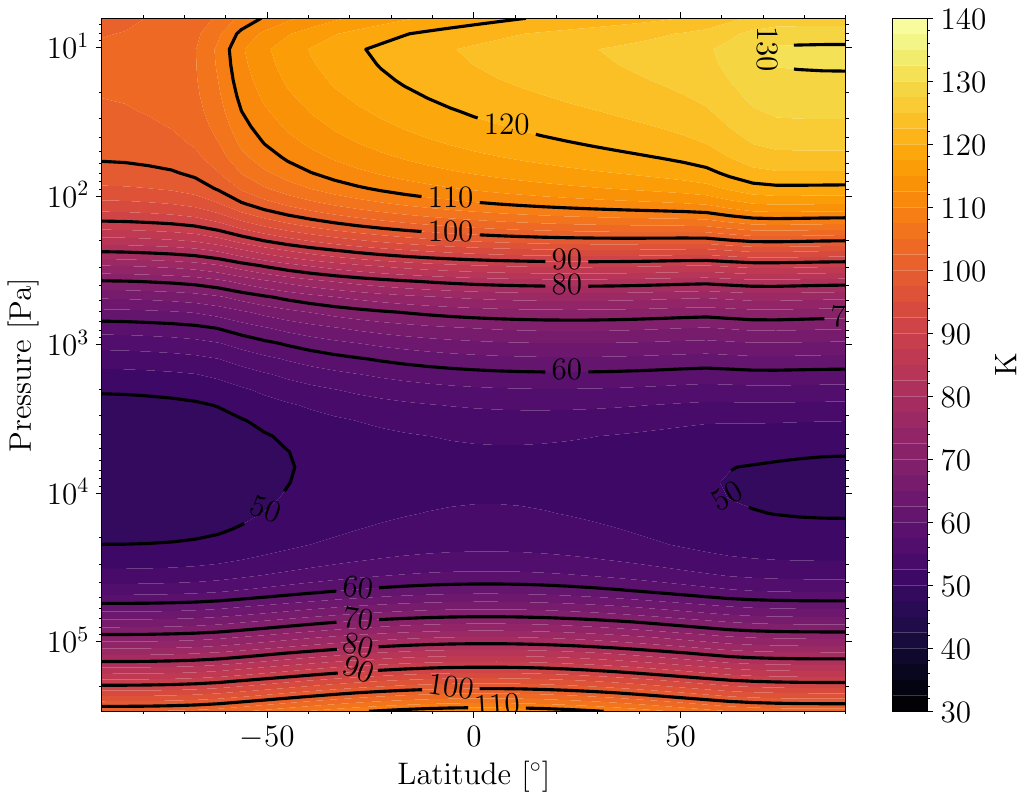}
  \label{fig:neptune-max}}
  \hfill
\caption{\label{fig:neptune-nodyn}Neptune: Vertical cross sections of temperature at spring equinox Ls=0° (a), northern summer solstice Ls=90° (b) and during the northern maximum stratospheric temperature contrast at Ls=140° (c). The temperature observed at autumn equinox, winter solstice and the maximum seasonal contrast in the other hemisphere is almost the same but the maximum/minimum temperature is reversed in latitude.}
\end{figure}

On Neptune, the annual thermal amplitude is found to be greater at high latitudes than at the equator, as on Uranus. The \citet{Wallace1984} and \citet{Bezard1990} models give similar results but the amplitudes are very low (1 to 2 K for the poles and <0.5 K at the equator at $\sim$500 hPa). This is consistent with the annual contrast observed on our model (1.1 K at the poles and <0.1 K at the equator) (fig.\ref{fig:neptune-amplitude-500hpa}). The maximum north-to-south asymmetry occurs after each solstice in our model like on the previous models introduced above. \citet{Greathouse2011} predicts temperature near the winter solstice (Ls$\simeq$275°) to be 10 K warmer at the south pole than at the equator in the upper stratosphere (0.12 hPa) and the meridional temperature gradient becomes smaller at deeper levels (2.1 hPa). These previous predictions are in agreement with our simulations (fig.\ref{fig:neptune-nodyn}), yet none reproduce observations.

\subsection{Comparison to observed temperature contrasts on Uranus \label{sec:seasonal-obs-uranus}}

On Uranus, very few spatially-resolved observations probing the lower stratosphere have been made, contrary to Neptune (see section \ref{sec:seasonal-obs-neptune}), due to the cold temperatures and hence a poor signal-to-noise ratio. VLT/VISIR mid-infrared images at 13 $\mu$m \citep{Roman2020}, sensitive to the pressure level of 25 Pa, revealed that the meridional temperature trend between 20°S and 70°N remained unchanged between 2009 (Ls$\simeq$7°) and 2018 (Ls$\simeq$43°). They report a minimum temperature centred at the equator and a maximum at mid-latitudes (40°S and 40-60°N). 
In addition, localised longitudinal temperature variations that may be the manifestation of meteorological activity were reported \citep{Rowe-Gurney2021}. 
The meridional temperature variations simulated by our model are inconsistent with the observed ones \citep{Roman2020} in the lower stratosphere. Our radiative equilibrium model predicts a maximum temperature at the south pole and a minimum at the north pole between 7° and 43° in solar longitude (fig.\ref{fig:uranus-ls}). Concerning the seasonal variations, we predict a temperature increase (resp. decrease) of 3 K at 25 Pa  at the north (resp. south) pole between 2009 and 2018, which is at odds with the lack of observed seasonal trends between these two dates \citep{Roman2020}.

\begin{figure}[h]
    \centering
    \includegraphics[width=.45\textheight]{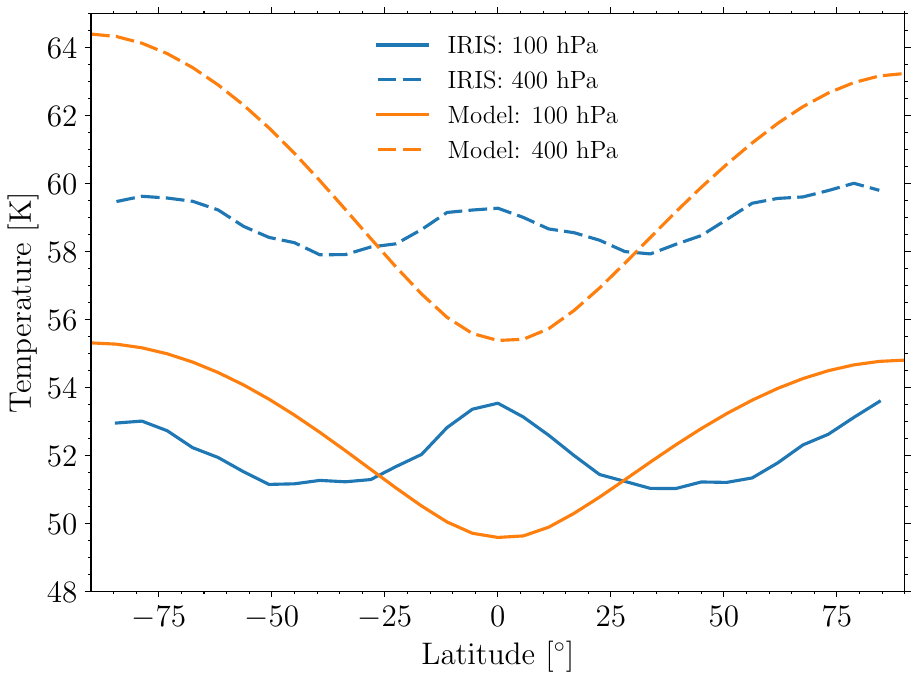}
    \caption{\label{fig:uranus-obs}Uranus: Comparison of the temperature retrieved (blue line) from Voyager~2/IRIS experiment \citep{Orton2015} and the simulated temperature (orange line) at 100 hPa (solid line) and 400 hPa (dashed line). Uncertainties from the spectral inversion of IRIS data are lower than 1~K. }
\end{figure}

The Voyager~2/IRIS experiment provides information about the thermal structure in the upper troposphere, between 70 and 400 hPa at Ls$\simeq$271° \citep{Conrath1998,Orton2015}. In this pressure range, derived temperatures by \citet{Orton2015} show minima at mid-latitudes (30–40°N and 20–50°S) and maxima at the equator and the poles. \citet{Conrath1998} show only one minimum located at 30°N. The source of the discrepancy between \citet{Conrath1998} and \citet{Orton2015} is unclear (see \citet{Orton2015} for discussion). The thermal structure predicted by our model is however different (fig.\ref{fig:uranus-obs}): the maximum is located at the poles and the minimum at the equator. However, the maxima/minima predicted at the tropopause are similar to those observed at higher levels, in the lower stratosphere \citep{Roman2020}.
Below the 100 hPa level, the temperature on IRIS data appears to be symmetric between the two hemispheres but above this level, a very slight asymmetry is present. At the tropopause (around 70 hPa), there is a temperature difference of 1 K between the two minima at mid-latitudes, with the northern (winter) latitudes exhibiting lower temperatures. 
This asymmetry could be explained by the eccentricity of Uranus where the perihelion occurs during the northern autumn equinox and IRIS observations were made during the northern winter solstice. However, the 1~K variation is probably within the uncertainties on the IRIS retrievals. A similar temperature anomaly is found in our simulations (fig.\ref{fig:uranus-ls}) but at higher latitudes and lower pressure. Thermal imaging performed one season after the IRIS observations (during the spring equinox in 2007) showed no significant change in tropospheric temperature between the two hemispheres. But observations after the spring equinox \citep{Roman2020} show a very slight increase in temperature in the northern/summer hemisphere (<0.3 K). A similar trend is observed in our simulations at the same period near the tropopause.

\subsection{Comparison to observed temperature contrasts on Neptune \label{sec:seasonal-obs-neptune}}

\begin{figure}[h]
    \centering
    \includegraphics[width=.45\textheight]{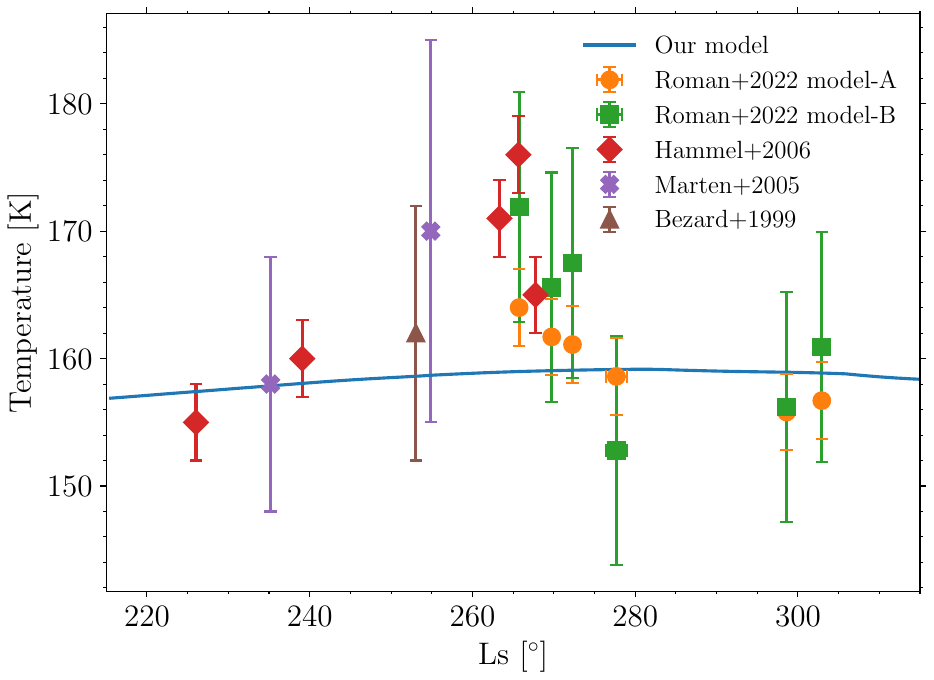}
    \caption{\label{fig:temporal-evolution}Comparison of the simulated disk-averaged temperature at 10 Pa on Neptune computed given the subsolar latitude observed from Earth (blue line) and the disk-averaged temperatures retrieved by \citet{Bezard1999} at $\sim$5 Pa (brown triangle), \citet{Marten2005} at $\sim$10 Pa (violet cross), \citet{Hammel2006} between $\sim$10 and 1 Pa (red diamond) and \citet{Roman2022} at $\sim$10 Pa (green square and orange circle). Due to the global offset in temperature between our simulations  and the observations (fig.\ref{fig:aerosols_conduction}), we have added +35K to our model results to help compare the trends in the datasets.}
\end{figure}

Contrary to Uranus, more measurements of the stratospheric temperature on Neptune have been made since the Voyager~2 era. Assuming a latitudinally-uniform distribution of methane as \citet{Greathouse2011}, \citet{Fletcher2014} compare thermal-infrared images from Keck/LWS (2003), Gemini-S/TReCS (2007) and built synthetic images from temperatures derived by IRIS on Voyager~2 (1989). A latitudinally-uniform temperature is retrieved both in 1989 (Ls$\simeq$235°) and 2007 (Ls$\simeq$275°), except at the south pole, where a warm polar vortex became evident only after the Voyager~2 flyby. However, stratospheric temperatures inferred between 2003 (Ls$\simeq$266°) and 2020 (Ls$\simeq$303°) show meridional variations between 0.1 and 0.5 hPa \citep{Roman2022}. In addition, significant temporal variations are highlighted by \citet{Roman2022}. Several temperature drops and increases of several kelvins (1 to 10 K) are observed during this period. Our simulated temperatures are inconsistent with the observed trends and meridional temperatures. At the solar longitudes of the observations (between 266° and 303°), a minimum is seen at the north pole and a maximum at the south pole as expected for a model with only pure radiative forcing at the southern summer. A warm south pole is observed since at least 2003 (Ls$\simeq$266°). Our simulations also predict this warm south pole where the maximum is reached between Ls$\simeq$272 and 333° at 0.1 hPa. The simulated temporal evolution of the temperature is much smoother than the observations (fig.\ref{fig:temporal-evolution}): the disk-averaged temperature calculated in our simulation taking the subsolar latitude into account shows a little variation ($\sim$2 K maximum) while the observed disk-averaged temperatures present important variations over shorter time scales at all latitudes and pressure levels. We note however that most observations agree with our predicted trends.  An exception concerns observations performed near Ls$\simeq$270° by \citet{Hammel2006} where the retrieved temperature significantly exceeds the general trend. 

At pressures lower than 0.1 hPa, a quasi-isothermal vertical structure with no temporal variation (less than 3 K) is observed from CH$_4$ emission \citep{Fletcher2014} but to be consistent with the temperature retrieved by \citet{Greathouse2011} at 0.007 hPa, this quasi-isothermal layer must be contained between 0.1 and 0.01 hPa. A similar quasi-isothermal vertical temperature is revealed by \citet{Roman2022} but an analogous meridional temperature variation as at deeper levels (0.1 hPa) is seen between 2003 and 2020. This vertically quasi-isothermal layer is also observed in our actual radiative-convective simulation (fig.\ref{fig:aerosols_conduction}) at pressures below 1 hPa.

\begin{figure}[h]
    \centering
    \includegraphics[width=.45\textheight]{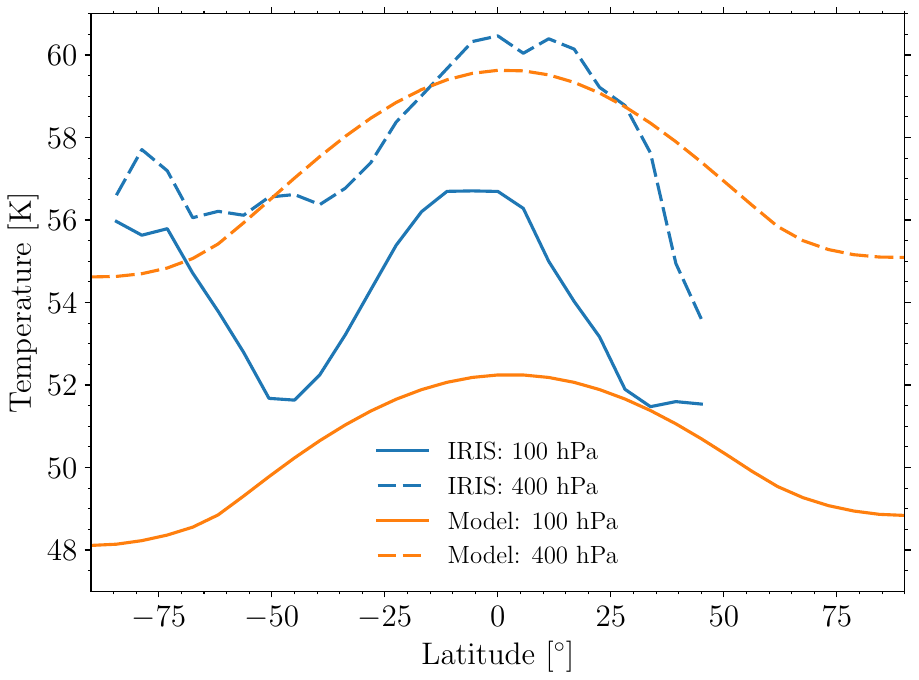}
    \caption{\label{fig:neptune-obs}Neptune: Comparison of the temperature retrieved (blue line) from IRIS experiment \citep{Fletcher2014} and the temperature simulated (orange line) at 100 hPa (solid line) and 400 hPa (dashed line). Uncertainties from the spectral inversion of IRIS data are lower than 1~K.}
\end{figure}

Concerning the upper troposphere, \citet{Fletcher2014} reanalysed IRIS data and retrieved the temperature between 70 and 800 hPa. A complex thermal structure has been revealed by this dataset which is rather similar to Uranus. At the tropopause (100 hPa), a minimum temperature of 51 K is present at mid-latitudes (+/- 45°) and a maximum temperature is reported at the south pole and the equator (56 K). The local minimum temperature in the northern hemisphere is slightly colder and appears to be higher in altitude than its southern counterpart. Due to the geometry of observation, no data beyond 40°N has been obtained. \citet{Roman2022} found that the thermal structure at the tropopause is quite different from that in the stratosphere, on average. The minimum temperature seen at mid-latitudes near the tropopause does not extend to the stratosphere.
In our simulation, the maximum temperature near the tropopause is also found at the equator but the south pole is colder and no temperature minimum is found at mid-latitudes (fig.\ref{fig:neptune-obs}). At 400 hPa, the retrieved temperature seems to be more consistent with our model in the southern hemisphere. 
Retrievals from mid-infrared spectroscopy observations performed with Keck/LWS in 2003 (Ls$\simeq$266°) \citep{Fletcher2014}, VLT/VISIR in 2006 (Ls$\simeq$272°), 2009 (Ls$\simeq$279°) and 2018 (Ls$\simeq$299°), Gemini-N/MICHELLE in 2007 (Ls$\simeq$270°) and Subaru/COMICS in 2008 (Ls$\simeq$277°) and 2020 (Ls$\simeq$303°) \citep{Roman2022} show that the temperature is unchanged at these levels since the Voyager~2 encounter, except for the south pole where the hot spot present at the upper troposphere and lower stratosphere (visible at 100 hPa in fig.\ref{fig:neptune-obs}) seems to be colder since 2018 but this cooling is uncertain. 
In our simulations, no change in temperature is observed at these pressure levels.

The thermal structure at the tropopause retrieved on Uranus and Neptune are very similar despite a different solar longitude of observations and seasonal forcings (most notably their obliquity). On both planets, the extrema near the tropopause are located at the same latitude band irrespective of the season. The mismatch between the observed thermal structure and the simulated one with our model suggests that the dynamics strongly influences the meridional temperature and is active over long time scales.
\citet{Flasar1987} and \citet{Conrath1987} proposed the presence of a mid-latitude upwelling and equatorial subsidence to explain the combination of low temperatures and sub-equilibrium para-fraction, while \citet{Karkoschka2009,Karkoschka2011} and \citet{Sromovsky2014} proposed an equatorial upwelling to explain cloud formation and the equator-to-pole methane gradient. 
From the combination of zonal wind, temperature, para-H$_2$ measurements and the distribution of condensable volatiles, \citet{dePater2014} and \citet{Fletcher2020} designed a model of the possible meridional circulation in the upper troposphere which reconciles the meridional temperature retrieved from IRIS observations and the meridional methane gradient. Above the methane condensation level, the proposed meridional circulation is ruled by a polar and equatorial subsidence and a mid-latitude upwelling. But below this level, the opposite is suggested by this model (with stacked circulation cells) which could explain the observed CH$_4$ vmr meridional gradient. The investigations of the meridional circulation on these planets remain to be confirmed with a 3-D general circulation model. 

In addition, the local minima visible at mid-latitudes may result from the thermal wind balance where if there is a decrease with height of the intensity of zonal jets \citep{Fletcher2014}, the meridional temperature gradients must be in balance with this decay.

If we take the latitudinal variation of methane observed in the troposphere of both planets \citep{Karkoschka2009,Karkoschka2011,Sromovsky2019,Irwin2019b} into account, the meridional temperature profile would be different but would still be far from the observations. On Uranus, this would imply a lower heating at the poles than at the equator, which would reduce or reverse the pole-equator contrast in our simulations. On Neptune, this contrast would be much more pronounced.

Photochemical models predict seasonal variations in hydrocarbon abundances in the stratosphere of ice giants \citep{Moses2018}. These variations could play an important role in heating and cooling in the polar regions, where variations of a factor of 10 are expected on Neptune for example.

\section{Summary and conclusions}

Radio-occultations and the IRIS experiment from Voyager~2 flyby provided important information about the thermal structure of Uranus and Neptune \citep{Lindal1987,Lindal1992,Conrath1998,Fletcher2014,Orton2015}. Given the low irradiance, the stratosphere of both planets is warmer than expected and this problem is called "energy crisis" (also affecting Uranus' and Neptune's thermospheres). Previous models had difficulty or failed to reproduce the observed stratospheric temperatures. One proposed solution was a higher (but unrealistic) concentration of stratospheric methane above the cold trap coupled with thermospheric conduction from an unknown source. Radiative forcing by aerosols was either not taken into account or if it was, it did not yield significant warming, due to incomplete observational constraints on the optical properties and physical characteristics of aerosol particles.

After the Voyager~2 era, the multiplication of near-infrared and thermal infrared observations has provided better constraints on the atmospheric composition and vertical aerosol/cloud structure of these planets and completed the temperature observations. The methane abundance on Uranus and Neptune retrieved by \citet{Lellouch2010,Lellouch2015} respectively are lower in the stratosphere than expected by older models. Moreover, the optical properties of hazes retrieved by several studies show that they are inconsistent with hydrocarbons known to condense on these planets (C$_2$H$_6$, C$_2$H$_2$, C$_4$H$_2$) according to photochemical models \citep{Moses2018,Moses2020,Dobrijevic2020}. The nature of the haze appears to be similar to tholins, with significant absorption in the near-infrared (> 1 $\mu$m) and more scattering in the visible, except near 300--400 nm.

In our study, these state-of-the-art observational constraints are used to better represent the radiative forcings in the atmospheres of Uranus and Neptune. The radiative-convective equilibrium model previously developed for gas giants \citep{Guerlet2014,Guerlet2020} is adapted here to the ice giants, firstly with only the contribution from the gaseous species. As expected, the radiative-convective model without additional heat sources produces the same results as previous models, i.e., an important gap between the simulated temperature profile and the observed one ($\sim$70 K on Uranus at 0.1 hPa and $\sim$25 K at 1 hPa on Neptune). 
To match the simulated profile to the observed one, several heat sources have been (re-)investigated. Altogether, our investigations allow us to write the following conclusions:

\begin{itemize}
    \item The haze scenario from \citet{Irwin2022} can reconcile the simulated temperature profile and the one observed by Voyager~2 on Uranus. The heating produced by the hazes is sufficient to reduce the previous gap of 70 K to 5-10 K at 0.1 hPa. Changing the refractive index retrieved by \citet{Irwin2022} to the one of tholins \citep{Khare1984} allows for a significant warming of the temperature profile, while warming by ice tholins is not sufficient \citep{Khare1993}. Very dark particles similar to dust particles falling from rings can also heat the stratosphere of Uranus but seem inconsistent with respect to the haze scenario from \citet{Irwin2022}. Concerning Neptune, the simulated temperature with the haze scenario from \citet{Irwin2022} becomes 10 K warmer than the case without aerosols. However, this is still too cold in comparison to the observed temperature profile. Unlike Uranus, no significant change has been obtained if the refractive index is replaced by that of tholins or ice tholins. With a hypothetical haze layer composed of very dark particles, the stratosphere of Neptune can be warm significantly.
    \item With a larger amount of stratospheric methane, it is possible to warm the temperature profile for both planets as was found in previous studies \citep{Appleby1986,Marley1999}. However, the methane abundance necessary to sufficiently warm the temperature profile is inconsistent with observations \citep{Lellouch2010,Lellouch2015}. A CH$_4$ vmr on the order of 10 times higher without hazes (or 4 times higher with hazes) on Uranus and 100 times higher on Neptune in the stratosphere is needed (with or without hazes).
    \item Thermospheric conduction cannot warm the stratosphere alone on both planets as confirming previous studies \citep{Wang1993,Marley1999}. Nevertheless, in combination with hazes or a larger amount of methane, conduction allows us to better match the observed temperature profile on Uranus in the upper levels of our model (for pressures below 50 Pa). On Neptune, the effect of conduction is hardly noticeable. The difficulty to warm the stratosphere of Neptune by radiative forcing or conduction leads us to hypothesise that the heating source may have a dynamical origin.
\end{itemize}

We note that the impact of haze and cloud layers on the heating rates has been investigated with the assumption of Mie scattering theory (spherical particles). However, a more complex structure such as fractal aggregate particles would be worth testing in future studies. Indeed, \citet{Zhang2015} showed that this type of particle dominates the heating rates in the stratosphere of Jupiter. We cannot yet test this hypothesis in the absence of retrieved haze properties assuming fractal aggregates.

Other heat sources not investigated in this study could also warm the stratosphere, such as 
heat released by inertia-gravity waves dissipation \citep{Roques1994}. 
Auroral heating could also 
heat the upper atmosphere of outer planets. \citet{Brown2020} and \citet{ODonoghue2021} show that for Saturn and Jupiter respectively, this process can inject a huge amount of energy about the magnetic poles. However, due to their dominant zonal circulation, the redistribution of heat towards lower latitudes remains difficult to explain. \citet{ODonoghue2021} find that planetary wave drag can advect the heat from the pole to the equator. The magnetic poles of Uranus are located at $\pm30^\circ$ in latitude and $\pm 43^\circ$ for Neptune. It would also be interesting to study the link with solar activity \citep{Roman2022}.

The IRIS experiment onboard Voyager~2 provided key information about the latitudinal thermal structure on these planets. Near the tropopause, the observed temperature is found minimal at mid-latitudes and maximal at the equator and poles on both planets. Surprisingly, despite different radiative forcing (owing to their obliquity and different solar longitude at the date of observation), their thermal structure is relatively similar. 
By analysing the temperature simulated in our model, two main conclusions can be drawn:

\begin{itemize}
    \item The simulated meridional thermal structure is inconsistent with the one observed by Voyager~2 at the same solar longitude on Uranus and Neptune and also by ground-based observations. The location of minima and maxima are at odds between our simulations and the observations. For Uranus, the simulated temperature near the tropopause is quite in agreement with the one retrieved from the IRIS experiment at high latitudes. However, the local temperature maximum at the equator is not produced by our model. On Neptune, the temperature at the tropopause is only consistent at the equator. The origin of the observed local maxima and minima could be explained by a meridional circulation which consists in subsidence at the equator and poles (adiabatic warming) and upwelling at mid-latitudes (adiabatic cooling) as proposed by \citet{Fletcher2020} and by thermal wind balance.
    \item In our simulations, a seasonal variation is seen above the tropopause where the maximum latitudinal contrast is shifted by Ls=+90° following the summer solstice on Uranus and +50° on Neptune. Below the tropopause, very slight variations are simulated on both planets, consistently with long radiative time scales. Contrary to simulations, no significant seasonal change has been observed since the Voyager~2 flyby at the tropopause and lower stratosphere on Uranus \citep{Roman2020}. On Neptune, significant variations in temperature have been observed since the Voyager~2 flyby \citep{Roman2022}. These variations cannot be explained by fully radiative seasonal forcings and suggest other effects with a subseasonal trend. This may indicate that the dynamics control the thermal structure at these levels, or that the assumption of a latitudinally-uniform methane abundance and aerosol distribution in the lower stratosphere is not appropriate.
\end{itemize}

Recent and future observations from JWST should provide new information on the thermal structure of Uranus and Neptune's troposphere and stratospheres with the MIRI (Mid-InfraRed Instrument) spectrometer and potentially on their haze physical properties and spatial distribution with NIRSPEC (Near-Infrared Spectrograph) and NIRCam (Near-InfraRed Camera) \citep{Norwood2016}.
First spectroscopic observations performed as part of a Guaranteed Time Observation program (PI: Leigh Fletcher) took place in January 2023 for Uranus and June 2023 for Neptune in the thermal infrared with MIRI.
At that time, Uranus was close to the summer solstice (Ls$\simeq$64°) and Neptune's season corresponds to southern summer (Ls$\simeq$310°). At 0.1 hPa, the thermal contrast predicted from our 1-D seasonal model on Uranus is 2.2 K between the equator and the north pole and 7.1 K for Neptune between the equator and the south pole. 
These new observations will enhance our knowledge of the thermal structure and its variations over time on these planets, and will help to constrain and complete future seasonal radiative-convective models. At the same time, the development of GCMs that take atmospheric dynamics into account appears necessary, given the puzzles that remain regarding the link between their thermal structure and meridional circulation.~\\

\section*{Acknowledgements}~\\

G. Milcareck, S. Guerlet and A. Spiga acknowledge funding from Agence Nationale de la Recherche (ANR) project SOUND, ANR-20-CE49-00009-01.
The authors acknowledge also Jan Vatant d'Ollone for his contribution at an early stage of this project.
T. Cavali\'e acknowledge funding from CNES.
Fletcher and Roman were supported by a European Research Council Consolidator Grant (under the European Union's Horizon 2020 research and innovation programme, grant agreement No 723890) at the University of Leicester.

\bibliographystyle{apalike}
\bibliography{main}

\end{document}